\documentclass[useAMS,usenatbib]{mn2e}
\usepackage{graphicx}\usepackage{epsfig}\usepackage{epsf}
\usepackage{amsmath}\usepackage{amssymb}\usepackage{stfloats}
\title[SN 2009ip was a core-collapse SN]{SN~2009\lowercase{ip} and
  SN~2010\lowercase{mc}: Core-collapse Type II\lowercase{n} supernovae
  arising from blue supergiants}

\author[Smith et al.]{Nathan Smith$^{1}$\thanks{E-mail:
    nathans@as.arizona.edu}, Jon C.\ Mauerhan$^{1}$, and Jose L.\ Prieto$^2$ \\
  $^{1}$Steward Observatory, University of Arizona, 933 N. Cherry
  Ave., Tucson, AZ 85721, USA \\ $^2$Department of Astrophysical
  Sciences, Peyton Hall, 4 Ivy Lane, Princeton, NJ 08540, USA}
\begin{document}

\pagerange{\pageref{firstpage}--\pageref{lastpage}} \pubyear{2012}
\maketitle
\label{firstpage}

\begin{abstract}

  The recent supernova (SN) known as SN~2009ip had dramatic precursor
  eruptions followed by an even brighter explosion in 2012.  Its
  pre-2012 observations make it the best documented SN progenitor in
  history, but have fueled debate about the nature of its 2012
  explosion --- whether it was a true SN or some type of violent
  non-terminal event.  Both could power shock interaction with
  circumstellar material (CSM), but only a core-collapse SN provides a
  self-consistent explanation.  The persistent broad emission lines in
  the spectrum require a relatively large ejecta mass, and a
  corresponding kinetic energy of at least 10$^{51}$~erg, while the
  faint 2012a event is consistent with published models of
  core-collapse SNe from compact ($\sim$60 $R_{\odot}$) blue
  supergiants.  The light curves of SN~2009ip and another Type~IIn,
  SN~2010mc, were nearly identical; we demonstrate that their spectra
  match as well, and that both are standard SNe~IIn.  Our observations
  contradict the recent claim that the late-time spectrum of SN~2009ip
  is returning to its progenitor's LBV-like state, and we show that
  late-time spectra of SN~2009ip closely resemble spectra of SN~1987A.
  Moreover, SN~2009ip's changing H$\alpha$ equivalent width after
  explosion matches behavior typically seen in core-collapse SNe~IIn.
  Several key facts about SN~2009ip and SN~2010mc argue strongly in
  favor of a core-collapse interpretation, and make a non-terminal
  10$^{50}$~erg event implausible.  The most straighforward and
  self-consistent interpretation is that SN~2009ip was an initially
  faint core-collapse explosion of a blue supergiant that produced
  about half as much $^{56}$Ni as SN~1987A, with most of the peak
  luminosity from CSM interaction.

\end{abstract}


\section{Introduction}

The violent episodes of mass loss that occur in the latest
evolutionary phases of some massive stars represent an important
unsolved problem in astrophysics.  The fits of eruptive and explosive
non-terminal mass loss help define the observational phenomenon
referred to collectively as luminous blue variables (LBVs); they have
no identified physical driving mechanism, despite the fact that they
may dominate the mass lost during the lives of the most massive stars
(Smith \& Owocki 2006).  Historically associated with the most
luminous blue supergiant stars in the Milky Way and its nearest
neighbors (Hubble \& Sandage 1953; Conti 1984; Humphreys \& Davidson
1994; Smith et al.\ 2004), LBVs have also been linked with a class of
extragalactic transient sources that have typical peak absolute visual
magnitudes around $-$14 mag, with outflow speeds of several hundred km
s$^{-1}$ (Smith et al.\ 2011a; Van Dyk et al.\ 2000; Kochanek et al.\
2012).  This diverse class of eruptions has long been discussed in the
context of extreme winds driven by super-Eddington luminosities (e.g.,
Owocki et al.\ 2004), but growing evidence suggests that some cases
can be explained as hydrodynamic (but non-terminal) explosions that
power their emergent radiation through a shock interacting with dense
circumstellar material (CSM), as in the prototypical case of $\eta$
Carinae (Smith 2013a; 2008).

Type IIn supernovae (SNe) may result from a special case of these
eruptions in the very latest phases of massive star evolution before
core collapse.  Their namesake narrow H lines indicate dense, slow CSM
surrounding the SN, which in turn requires eruptive LBV-like mass loss
to eject so much mass so soon before core collapse.  Evidence
providing a strong link between LBVs and SNe IIn comes in two key
flavors: (1) Super-luminous SNe~IIn, where the demands on the amount
of H-rich CSM mass are so extreme (10-20 $M_{\odot}$ in some cases)
that very massive stars are required, and the inferred radii and
expansion speeds of the CSM require that it be ejected in a short time
within just a few years before core collapse (Smith et al.\ 2007,
2008, 2010a; Smith \& McCray 2007; Woosley et al.\ 2007; van Marle et
al.\ 2009), and (2) direct detections of progenitors of SNe~IIn that
are consistent with massive LBV-like stars (Gal-Yam \& Leonard 2009;
Gal-Yam et al.\ 2007; Smith et al.\ 2010b, 2011a, 2012; Kochanek et
al.\ 2011).  We should note, however, that not all SNe IIn are
necessarily tied to LBVs and the most massive stars.  Some SNe~IIn may
actually be Type Ia explosions with dense CSM (e.g., Silverman et al.\
2013 and references therein), some may be electron-capture SN
explosions of stars with initial masses around 8--10 $M_{\odot}$
(Smith 2013b; Mauerhan et al.\ 2013b; Chugai et al.\ 2004), and some
may arise from extreme red supergiants like VY~CMa with very dense
winds (Smith et al.\ 2009a, 2009b; Mauerhan \& Smith 2012; Chugai \&
Danziger 1994).

%%%%%%%%%%%%%

SN~2009ip was nearby, and an extraordinarily well-observed event.  The
progression of the several-year pre-SN variability that culminated in
the dramatic 2012 event has now been recounted in detail by several
authors (Smith et al.\ 2010b, 2013; Foley et al.\ 2011; Mauerhan et
al.\ 2013a; Prieto et al.\ 2013; Pastorello et al.\ 2013; Fraser et
al.\ 2013a; Margutti et al.\ 2013).  First discovered in 2009 by the
CHASE project (Maza et al.\ 2009) and originally given a SN
designation, the object turned out to be an LBV-like eruptive
transient (at least until 2012).  SN~2009ip displayed a series of
brief brightening episodes with absolute magnitude peaks of roughly
$-$14 mag, consistent with LBV-like eruptions (Smith et al.\ 2011a),
as well as a longer duration S Dor-like eruption before that (Smith et
al.\ 2010b).  Spectra of these outbursts exhibited narrow emission
lines indicating dominant outflow speeds around 600 km s$^{-1}$,
although a small amount of material was accelerated to higher speeds
of several thousand km s$^{-1}$ (Smith et al.\ 2010b; Foley et al.\
2011; Pastorello et al.\ 2013).  This is reminiscent of $\eta$ Carinae
in its 19th century eruption (Smith 2006, 2008, 2013a). 

\begin{figure*}
\includegraphics[width=6.9in]{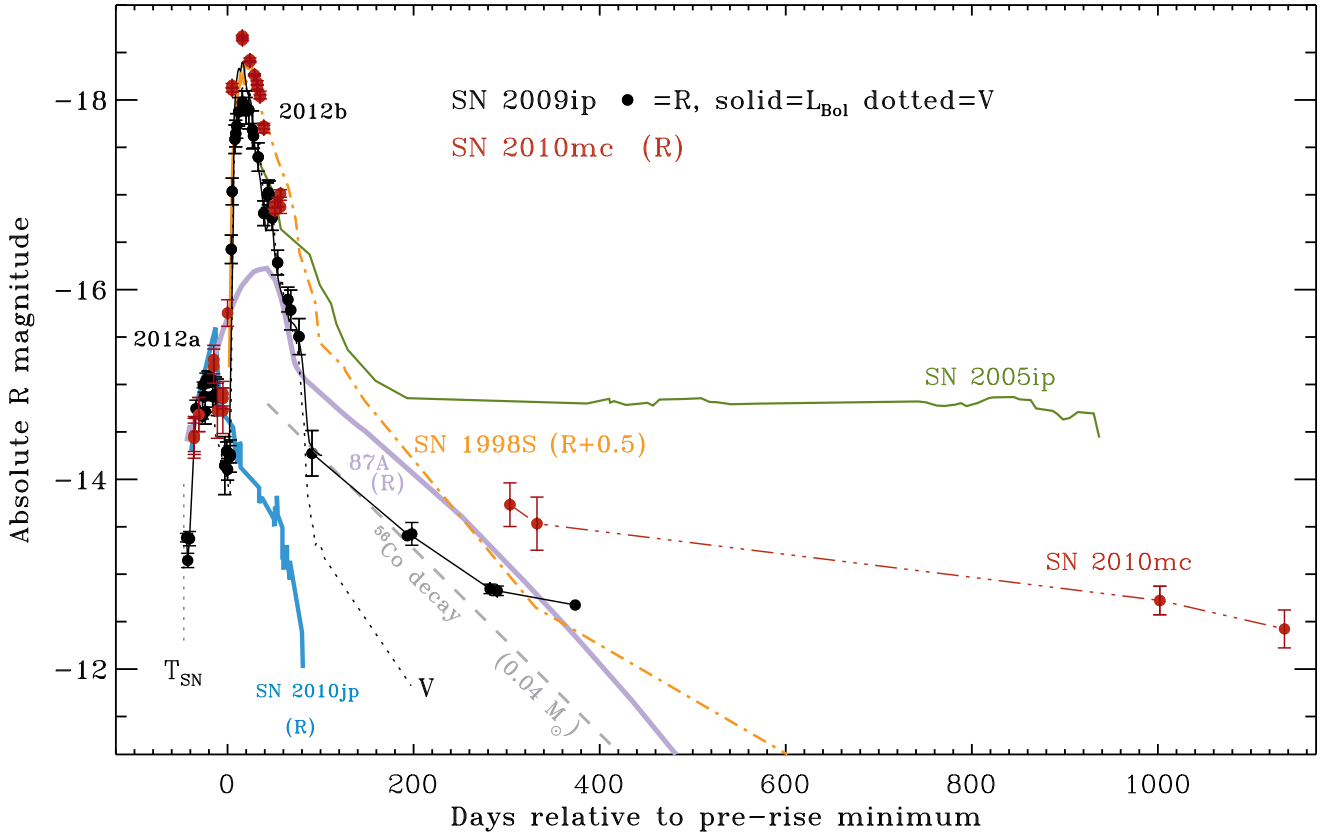}
\caption{Absolute-magnitude $R$-band light curves of SN~2009ip (black
  filled dots) and SN~2010mc (red filled dots).  This is an updated
  version of the figure in Smith et al.\ (2013), comparing the two SNe
  with no shift applied to either object (except for the shift
  necessary to convert the AB magnitudes used by Ofek et al.\ to Vega
  magnitudes used here).  We have now added late-time photometry
  points from Ofek et al. (2013a) for SN~2010mc, as well as our own
  late-time $r^\prime$ measurements after day 1000).  We include our
  own measurement of the late-time $R$-band magnitudes of SN~2009ip,
  as well as the preliminary reported magnitude given by Fraser et
  al.\ (2013b).  The solid black line is the bolometric luminosity of
  SN~2009ip, and the dotted black line is the $V$ magnitude, both from
  Margutti et al. (2013). Day = 0 is set to be the beginning of the
  sharp rise to peak of each object, after the precursor outburst.
  The gray dashed line represents radioactive decay luminosity from
  $^{56}$Co for a synthesized $^{56}$Ni mass of 0.04 $M_{\odot}$,
  assuming that the time of the SN explosion ($T_{SN}$) is at $-$47
  days, corresponding to just before the rise of the 2012a peak.  For
  comparison, we also show (in blue) the $R$/unfiltered light curve of
  SN~2010jp from Smith et al.\ (2012), the $R$-band light curve of
  SN~2005ip (in green; Smith et al.\ 2009b), and the $R$-band light
  curve of SN~1998S (yellow dot-dash; Fassia et al.\ 2000; Poon et
  al.\ 2011; including late-time measurements of the decline rate from
  Li et al.\ 2002).  No shift was applied to SN~2010jp or SN~2005ip,
  but SN~1998S is scaled fainter by $+$0.5 mag for comparison. The
  $R$-band light curve of SN~1987A (thick lavender curve) is also
  shown (Hamuy et al.\ 1990).}
\label{fig:phot}
\end{figure*}

SN~2009ip then sufferred a much more extreme event in 2012 (discovered
by Drake et al.\ 2012), when it first brightened to roughly $-$15
mag and exhibited broad emission lines with outflow speeds of order
$\sim$10,000 km s$^{-1}$ (Mauerhan et al.\ 2013a).  This first peak in
August/September (2012a) initially faded by $\sim$0.8 mag over 20
days, but was followed by an extremely rapid rise to a second peak in
October (2012b) with a luminosity of roughly $-$18.0 mag ($R$-band;
see Figure~\ref{fig:phot}).  Mauerhan et al.\ (2013a) proposed that
the initial fainter 2012a peak was the actual SN explosion, whereas
the fast rise to peak in 2012b was caused by CSM interaction as the
fast SN ejecta caught up to the CSM ejected in the LBV-like eruptions
one or more years before.

If the 2012 event really was a core-collapse SN, it would dramatically
confirm that LBV-like stars do in fact explode as SNe, which would
expose significant errors in the most fundamental paradigms of current
stellar evolution models.  As reviewed by Langer (2012), models for
massive star evolution envision that very massive stars in the local
universe will shed all their H envelopes via stellar winds, will
become Wolf-Rayet stars, and will explode as H-poor SNe of Types Ib or
Ic (see e.g., Heger et al.\ 2003; Maeder \& Meynet 2000; Langer et
al.\ 1994).  Thus, the case of SN~2009ip - with a very massive
LBV-like star exploding as a SN~IIn before losing its H envelope ---
would be a very important discrepancy with models of massive star
evolution.  Smith \& Arnett (2013) have recently discussed a likely
reason why 1-D stellar evolution calculations would fail to account
for the pre-SN variability that is observed, centered on the treatment
of convection in late burning stages.

%%%%%

However, the idea that the 2012 brightening of SN~2009ip was a true
core-collapse SN has been controversial.  The initial faintness of the
2012a outburst caused some doubt (Margutti et al.\ 2012a; Martin et
al.\ 2012), but some core-collapse SNe are faint initially if the
progenitor has a compact radius as in the vivid case of SN~1987A
(e.g., Arnett 1989).  In any case, SN~2009ip then brightened very
quickly to a luminosity commensurate with normal SNe (Brimacombe 2012;
Margutti et al. 2012b; Smith \& Mauerhan 2012b).  Pastorello et
al. (2013) proposed that the 2012 event of SN~2009ip was not a SN,
referring to the broad absorption wings seen in the precursor events
in previous years as evidence that high speeds do not necessarily
require a final core-collapse event.  Pastorello et al.\ (2013)
instead favored the idea that the 2012 event of SN~2009ip was the
result of a non-terminal pulsational pair instability (PPI) eruption
(e.g., Heger \& Woosley 2002; Woosley et al.\ 2007; Chatzopoulos \&
Wheeler 2012).  Fraser et al.\ (2013a) presented additional data and
continued to favor the PPI model instead of core collapse, based on
the lack of nebular features during the decline that are seen seen in
normal SNe, and they predicted that the star would therefore survive
and return to its previous state when it emerged from behind the Sun.
Margutti et al.\ (2013) presented a very extensive and comprehensive
analysis of multiwavelength spectra and photometry during the main
2012 bright phase of SN~2009ip.  These data were consistent with the
idea that the main properties of SN~2009ip could be explained by
modest CSM interaction involving a total energy of at least 10$^{50}$
erg, but these authors favored the interpretation of a non-terminal
shell ejection event.

%%%%%

The crux of the debate is that it is not immediately obvious that
SN~2009ip was a core-collapse event, since it is a Type IIn supernova
--- these are strongly influenced by CSM interaction, and so they
don't always look like ``normal'' examples of familiar SNe with
well-defined photospheric and nebular phases.  In particular, the
nucleosynthetic products normally seen in nebular features are usually
masked in SNe~IIn due to ongoing CSM interaction.  Moreover, since CSM
interaction is the main engine, any type of fast ejecta crashing into
a slower CSM shell might power a luminous display (e.g., Dessart et
al.\ 2009; Woosley et al.\ 2007; Smith 2013b).  Due to the high
efficiency of converting kinetic energy into radiated luminosity (van
Marle et al.\ 2009; Smith \& McCray 2007; Woosley et al.\ 2007), even
relatively low-energy ($\sim$10$^{50}$ erg) explosions can in
principle produce bright transients comparable to SNe if most of the
kinetic energy is tapped.  Such low energy explosions combined with
CSM interaction have recently been proposed as viable power sources
for the Crab's SN in 1054 AD (Smith 2013b), $\eta$ Car's 19th century
eruption (Smith 2013a), and SN~1994W (Dessart et al.\ 2009).

There are, however, clues in the spectroscopic evolution that must be
reconciled with the energy budget in CSM interaction, and there are
potentially some clear tests in the late-time nature of the surviving
(or not) object.  When SN~2009ip re-emerged from behind the Sun in
2013, Fraser et al.\ (2013b) reported a spectrum and photometry, and
claimed that the star was returning to its pre-SN LBV state.  Our
late-time data reported here strongly contradict this conclusion, as
detailed below.  Moreover, we argue that observations require the
combination of a normal core-collapse SN energy
($\sim$1$\times$10$^{51}$ ergs) combined with CSM interaction in an
asymmetric environment.

The present paper is organized around three main points, which serve
to argue that 2009ip was indeed a core-collapse SN as originally
proposed by Mauerhan et al.\ (2013a):

1. We argue that SN~2009ip (2012a/b) was essentially identical to
SN~2010mc.  Smith et al.\ (2013) already showed this based on
photometry, which was also found in the analysis by Margutti et al.\
(2013).  Here we extend the comparison to spectra of SN~2010mc and
SN~2009ip, showing that they are carbon copies of one another at early
times.

2. We present new late-time photometry and spectra of both SN~2009ip
and SN~2010mc, both of which appear consistent with Type~IIn
supernovae dominated by CSM interaction at late times, but which do
not indicate a return to the LBV-like states of the
progenitors. SN~2009ip, in particular, shows the trend of strongly
increasing H$\alpha$ emission equivalent width, which in SNe IIn is
due to the continuum optical depth decreasing when the CSM shell
expands and becomes transparent.  After 3 years, SN~2010mc has only
continued to fade, showing no sign of renewed LBV-like activity, and
its late time spectrum shows no detectable continuum and only a faint
and narrow H$\alpha$ line consistent with late-time CSM interaction in
a decelerated shell.  Contrary to claims made by Fraser et al.\
(2013b), our late time spectrum of SN~2009ip clearly does not resemble
its previous LBV-like state.  Instead, it appears very similar to
spectra of well studied core-collapse SNe like SN~1999em and SN~1987A,
with the exception that roughly half the continuum luminosity and the
strong narrow emission lines come from CSM interaction luminosity.  
% At the latest epochs, SN~2009ip and SN~2010mc appear to resemble a
% late-time (day 445) spectrum of the super-luminous SN~IIn
% 2006tf. This comparison is valuable because SN~2006tf emitted almost
% 10$^{51}$ ergs in visual luminosity, and was almost certainly a real
% SN.

3.  We point out several other aspects relevant to this discussion
that make a non-SN interpretation physically unlikely, statistically
improbable, and not self consistent.  In particular, when one takes
into account an asymmetric CSM, this has important implications for
the total energy budget --- if CSM interaction occurs in only a small
fraction of the solid angle, then the mass and explosion energy
derived from CSM-interaction diagnostics are only lower limits.
Moreover, we show that the explosion must have been substantially
asymmetric to account for the persistent broad lines seen at late
times, and the SN ejecta must have been substantially more massive
than the minimum required for CSM interaction because they remain
opaque for several months. This, combined with the fast expansion
speeds observed in spectra, raises the required explosion energy to be
at least 10$^{51}$ ergs.  Finally, we demonstrate that the light curve
can be explained quite well with published models of core-collapse SNe
from a blue supergiant, like SN~1987A, with the caveat that the 2012b
peak was dominated by CSM interaction as indicated by spectra.

\section{OBSERVATIONS}

%%%%%%%%%%%%%%%%%%%%%%%% TABLE %%%%%%%%%%%%%%%%%%%%%%%%%%%%%%%%%%%%%%%
%\begin{center}
\begin{table}\begin{center}\begin{minipage}{3.3in}
      \caption{Photometry of SN~2009ip and 2010mc}
\scriptsize
\begin{tabular}{@{}llllccc}\hline\hline
SN  &UT date &day$^a$ &Tel./Instr.  &filt. &mag &err \\ \hline
09ip  &2013 Apr 6/7      &195     &Mag./IMACS   &$R$   &18.19   &0.12  \\
09ip  &2013 Apr 5-7      &194     &d.P./WFCCD    &$R$    &18.18   &0.12  \\
09ip  &2013 Jun 30       &282     &d.P./WFCCD    &$R$     &18.76   &0.05  \\
09ip  &2013 Jul 8        &290     &Mag./IMACS   &$R$      &18.78   &0.05  \\
09ip  &2013 Aug 30       &344     &d.P./WFCCD    &$R$        &18.76   &0.05  \\
09ip  &2013 Sep 30       &374     &LBT/MODS         &$r^{\prime}$ &18.93   &0.02  \\
10mc  &2013 May 18       &1002    &LBT/MODS         &$r^{\prime}$ &23.45   &0.15  \\
10mc  &2013 Sep 30       &1136    &LBT/MODS         &$r^{\prime}$ &23.75   &0.2   \\
\hline
\end{tabular}\label{tab:p48}\end{minipage}
\end{center}
$^a$For SN~2009ip and 2010mc, we define day=0 as the time of the
onset of the rapid rise to peak luminosity.  As discussed in the text,
we adopt an explosion date 47 days before this ($T_{SN}$ = $-$47 d).
\end{table}%\end{center}

\subsection{Photometry}

The uncanny similarity between the light curves of SN~2009ip and
SN~2010mc was first noted by Smith et al.\ (2013), and
Figure~\ref{fig:phot} is an updated version of the plot in that paper,
but with an extended range of time.  Margutti et al.\ (2013) also
discussed the similarity between the light curves of these two
events.\footnote{Note that the peaks of SN~2009ip and SN~2010mc are
  closer to one another in Margutti et al.'s plot than here in
  Figure~\ref{fig:phot}. This is because Margutti et al.\ adopted a
  larger distance to SN~2009ip of 24 Mpc, as compared to 20.4 Mpc
  adopted here.  The SN~2009ip photometry is in good agreement in
  various studies.}  The early ($t<100$ days) photometry is comprised
of published measurements (Mauerhan et al.\ 2013a; Prieto et al.\
2013; Ofek et al.\ 2013a; Pastorello et al.\ 2013), as noted
previously by Smith et al.\ (2013).  The later measurements are a
combination of new data presented here and in recent reports.  We plot
absolute magnitudes in Figure~\ref{fig:phot}, adopting a distance
modulus of 31.55 mag for SN~2009ip's host galaxy NGC 7259 and an
extinction value of $A_R$=0.051 mag (Smith et al.\ 2010b).

After it reappeared from behind the Sun, we obtained imaging
photometry of SN~2009ip at several epochs using various facilities,
summarized in Table 1.  We obtained $R$-band photometry using IMACS in
imaging mode on the 6.5 m Baade telescope at the Magellan Observatory,
as well as the du Pont 2.5 m telescope at Las Campanas, and
$r^{\prime}$ photometry using the imaging acquisition mode of the MODS
spectrograph on the Large Binocular Telescope (LBT).  All nights were
photometric, and SN~2009ip was bright enough and isolated enough that
aperture photometry was reliable.  Note that for the $R$ photometry
using the du Pont 2.5 m telescope, we obtained individual measurements
on three separate nights, April 5, 6, and 7 (UT), for which we derive
$R$ magnitudes of 18.19, 18.18, and 18.14 ($\pm$0.05 mag for each),
respectively.  The average of these measurements is $R$=18.18
($\pm$0.12), given in Table 1.

Figure~\ref{fig:phot} also includes the late-time measurement of
SN~2009ip in the preliminary report by Fraser et al.\ (2013), which
included no information about the uncertainty (so no error bar is
shown in Figure~\ref{fig:phot}).  They reported $R$=18.2 mag on 2013
Apr.\ 2; this was taken 5 days before our Magellan/IMACS measurement
in the same filter, and agrees very well with ours to within our
uncertainty.

\begin{figure}
\includegraphics[width=3.0in]{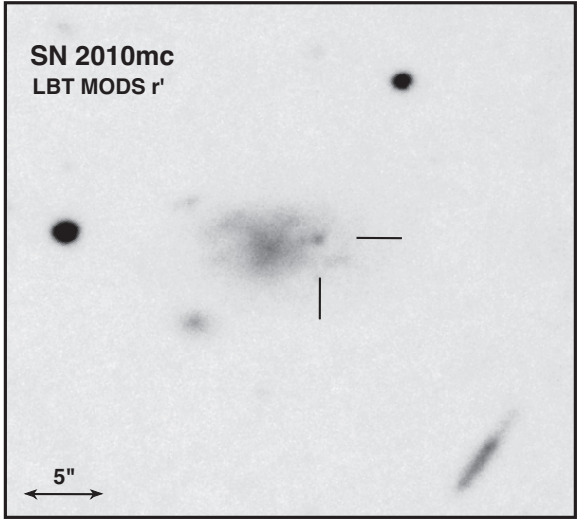}
\caption{An LBT/MODS $r^\prime$ image of SN~2010mc and its host
  galaxy, taken in April 2013 (day 1002 in Figure~\ref{fig:phot}).}
\label{fig:img}
\end{figure}

We also observed SN~2010mc at late times using LBT/MODS, but here
photometry was more difficult and influenced by seeing, since
SN~2010mc was much fainter and embeded within its host dwarf galaxy
light (Figure~\ref{fig:img}).  For this reason, late-time measurements
for SN~2010mc have relatively large error bars.

\begin{figure}
\includegraphics[width=3.3in]{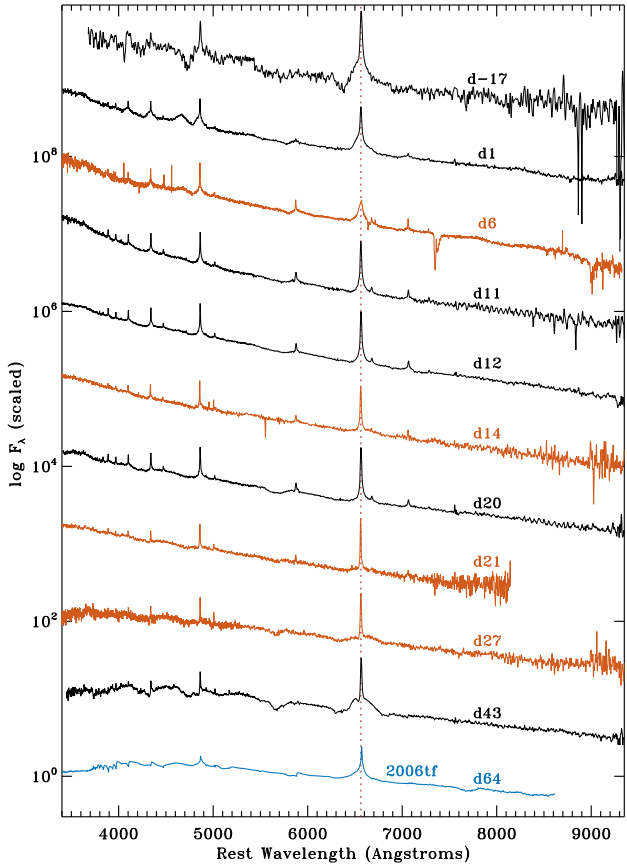}
\caption{Early-time spectral evolution of SN~2009ip (black) and
  SN~2010mc (orange).  All spectra here have been previously
  published, but have not been compared to one another.  SN~2009ip
  spectra are from Mauerhan et al.\ (2013a), SN~2010mc spectra are
  from Ofek et al.\ (2013a), and the spectrum of SN~2006tf (blue) is
  from Smith et al.\ (2008).}
\label{fig:earlyspec}
\end{figure}

\subsection{Archival Spectra of SN~2009ip and SN~2010mc}

Figure~\ref{fig:earlyspec} provides a comparison of the early spectral
evolution of SN~2009ip and SN~2010mc, during the main bright phases of
their outbursts, interleaved and ordered in time relative to the onset
of the rapid rise to peak (taken as 2012 Sep. 24 for SN~2009ip, and
2010 Aug.\ 20 for SN~2010mc).  These spectra were published by
Mauerhan et al.\ (2013a) and Ofek et al.\ (2013a), respectively, and
the Ofek et al.\ spectra were obtained from the publically available
WISeREP repository (Yaron \& Gal-Yam
2012)\footnote{http://www.weizmann.ac.il/astrophysics/wiserep/}.  For
reference, we also include a 2012a spectrum of SN~2009ip (day $-$17),
from Mauerhan et al.\ (2013a).  Figure~\ref{fig:earlyspec} shows the
day 64 (time after discovery) Keck spectrum of the super-luminous Type
IIn SN~2006tf for comparison (plotted in blue), published by Smith et
al.\ (2008).

\begin{figure}
\includegraphics[width=3.3in]{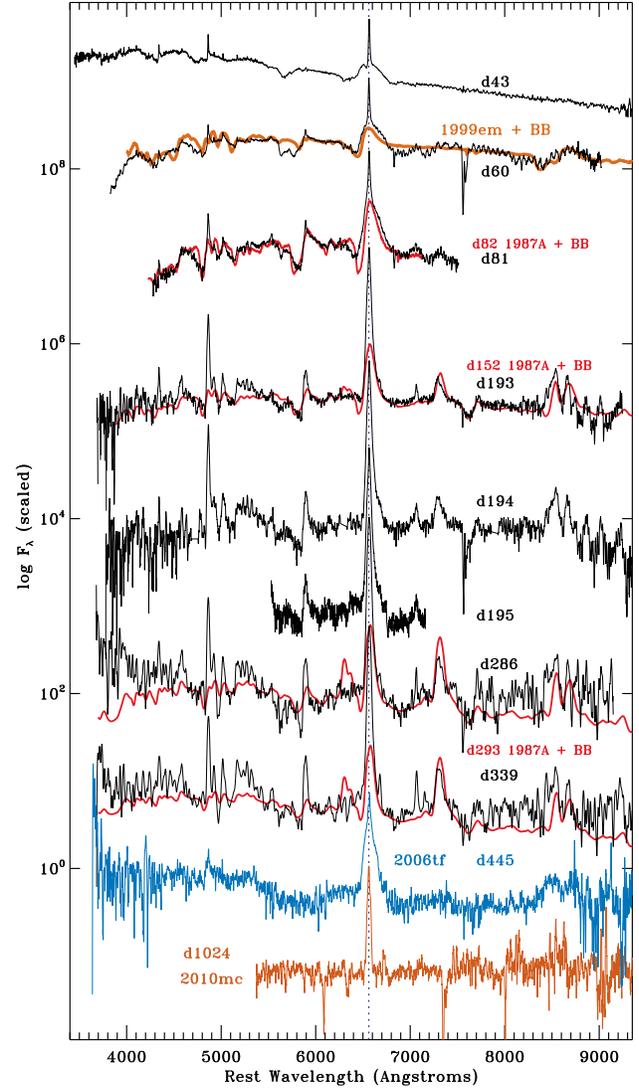}
\caption{Late-time spectral evolution of SN~2009ip after it faded from
  maximum luminosity, and more recently after it reappeared from
  behind the Sun. The first few spectra of SN~2009ip are reproduced
  from Mauerhan et al.\ (2013a).  The day 193-195 (listed as day 193),
  day 194, day 286, and day 339 spectra of SN~2009ip are from Las
  Campanas, the day 195 spectrum of SN~2009ip is from the MMT, the day
  1024 spectrum of SN~2010mc (orange) was obtained at the LBT, and the
  day 445 spectrum of SN~2006tf (blue) is from Smith et al.\
  (2008). For comparison with the day 60 spectrum of SN~2009ip, we
  show the observed spectrum of the normal Type II-P SN~1999em
  (orange) combined with a blackbody (50\% each) to veil the strength
  of spectral lines as described in Smith et al.\ (2010a).  Similarly,
  the day 81, 193, 286, and 339 spectra of SN~2009ip are compared to
  spectra of SN~1987A (red) at similar epochs obtained from the
  SUSPECT online database; as before, half the red continuum is from a
  blackbody.}
\label{fig:l8spec}
\end{figure}

\subsection{New Late-Time Spectra}

When SN~2009ip re-emerged from behind the Sun in April 2013, we
obtained visual-wavelength spectra to study its late-time behavior
after its decline from maximum luminosity.  We obtained spectra using
the WFCCD instrument on the du Pont 2.5-m telescope of the Las
Campanas Observatory, Chile. Obtained during clear conditions with the
slit oriented at the parallactic angle, this spectrum used a
1$\farcs$7 wide long-slit aperture and the 400 lines mm$^{-1}$
blue-sensitive grism, yielding $\sim$7 \AA \ resolution covering
3700-9300 \AA.  Spectra of SN~2009ip were obtained on the four
sequential nights of 2013 April 4$-$7, and these four spectra were
co-added to produce a single spectrum with higher signal-to-noise
ratio.  We refer to this combined spectrum as the day 193 spectrum,
obtained on days 193-196 relative to the pre-rise minimum.  We
obtained additional WFCCD spectra on 2013 July 8 (approximately day
286) and 2013 Aug 30 (day 339).

We observed SN~2009ip in spectroscopic mode using IMACS at Magellan.
On 2013 April 5 we obtained a low-resolution spectrum covering a wide
wavelength range from 3800--9520 \AA, using the 300 lines mm$^{-1}$
grating and a 1$\farcs$0 wide long-slit aperture oriented at the
parallactic angle.  The next night, 2013 April 6, we observed it again
with the same slit, but this time using the moderate-resolution 1200
lines mm$^{-1}$ grating covering 5560--7200 \AA.  In both spectra, CCD
chip gaps on IMACS yielded small sections of the spectrum with
interrupted wavelength coverage.  The dates for these two spectra
correspond to days 194 and 195 after the beginning of SN~2009ip's
sharp rise to maximum.  Both nights were photometric with good
0$\farcs$6 seeing.  An additional IMACS spectrum was obtained on 2013
June 30 (day 278).  This was obtained with a different configuration
using the f/2 camera and the 200 lpm grism with a 0$\farcs$9 slit
width at the parallactic angle, yielding spectra coverage from
4350--9800 \AA \ (with a small chip gap at 6408-6540 \AA) with a
resolution of $\sim$5 \AA.  The June 30 IMACS spectrum is similar to
other spectra that have higher signal to noise ratio, and so it is not
shown here, but is used to measure the H$\alpha$ equivalent width.  We
also observed SN~2009ip with the Bluechannel spectrograph at the MMT
on 2013 June 3, using a 1200 lines mm$^{-1}$ grating with wavelength
coverage of 5690--7000 \AA.  This last MMT spectrum is also not shown
in Figure~\ref{fig:l8spec} because it is very similar to the
Magellan/IMACS spectra, but it is used to measure the H$\alpha$
equivalent width, as discussed later in the paper.

We also obtained late-time spectra of SN~2010mc on 2013 Apr.\ 11 and
June 11 using the Multi-Object Double Spectrograph (MODS; Byard \&
O’Brien 2000) on the LBT.  Spectral images in the blue and red
channels were obtained with MODS in longslit mode, utilizing the G670L
grating and a 1.2$\arcsec$ slit, which yielded a spectral resolution
of R$\approx$1000.  A spectrum of SN~2010mc was also obtained on 2013
April 11 at the MMT using the bluechannel spectrograph, with higher
spectral resolution and a smaller wavelength range.  This spectrum has
a lower signal-to-noise ratio than our LBT spectra, but it confirms
the detection of broadened H$\alpha$ with very weak continuum.  The
2013 June 11 (day 1024) MODS spectrum is shown in
Figure~\ref{fig:l8spec}.  Finally, we observed both SN~2010mc and
SN~2009ip again with LBT/MODS on 2013 Sep 30, although on this night
only the red channel was functioning.  The spectra are not shown in
figures, but we use them to measure the H$\alpha$ equivalent width, as
discussed below.

In all cases, the spectral images were bias subtracted, flat fielded,
and median combined with a suitable rejection filter to remove cosmic
rays. The spectra were extracted using standard IRAF routines. Removal
of the background line emission from the sky and from the host galaxy
was accomplished by sampling the background from a narrow strip very
close to the dispersion track of the SN emission, such that the
background region overlapped very slightly with the wings of the SN
point spread function (PSF), but the result subtracted less than a few
per cent of the SN flux.  The extracted spectrum was wavelength
calibrated using spectra of HeNeAr lamps. The wavelength solutions
were corrected for the redshift of the host galaxies of SN~2009ip
(z=0.00572; Smith et al.\ 2010b) and SN~2010mc (z=0.035; Ofek et al.\
2013a).  The resulting late-time spectra are displayed in
Figure~\ref{fig:l8spec}, where we also include a previously published
late-time spectrum of SN~2006tf (Smith et al.\ 2008) for comparison.

Lastly, we obtained two late-time epochs of near-infrared (IR) spectra
of SN~2009ip using FIRE on the Magellan Baade telescope.  These were
obtained in low-resolution (prism) mode on 2013 April 17 (day 204) and
June 30 (day 278).  The data reduction was completed using the
standard FIREHOSE pipeline, in the same way as for our previous epochs
of FIRE spectra (see Smith et al.\ 2013 for details).  We show the day
278 near-IR spectrum in Figure~\ref{fig:irspec}, while the day 204
spectrum looks very similar and is not shown here.

\section{SN~2009\lowercase{ip} AND SN~2010\lowercase{mc} BOTH  LOOK LIKE
  NORMAL TYPE II\lowercase{n} SUPERNOVAE}

Here we aim to demonstrate that the time evolution of SN~2009ip is
entirely consistent with known examples of SNe~IIn, and that in
particular it is nearly identical to SN~2010mc.  This appears to be
true in terms of both its photometric and its spectroscopic
evolution. From around the time of peak luminosity onward, SN~2009ip
exhibits no behavior that diverges from observed properties of
``normal'' (i.e. moderate-luminosity) SNe~IIn.  What makes SN~2009ip
so unusual is that we have extensive and detailed pre-SN information
about the unstable progenitor, which might be present in the larger
sample of SN IIn, but simply goes undetected because of poorer
sensitivity in more distant SNe or inadequate time sampling.
%Recall that the CSM interaction present in {\it all} SNe~IIn requires
%rather extreme episodic, eruptive, or explosive pre-SN mass loss.  The
%hypothesis that SN~2009ip may actually be representative of a larger
%class of SNe~IIn sets the stage for the discussion in the last section
%(Section 4).

\subsection{Light Curves}

In a previous paper (Smith et al.\ 2013), we first pointed out that
the light curve of SN~2009ip's 2012 explosion is nearly identical to
that of SN~2010mc (Ofek et al.\ 2013a), without shifting the absolute
magnitude scale.  In particular, the timescales of the luminosity
evolution are astonishingly similar.  This is a strong indication that
SN~2009ip is not alone, and that whatever explanation we arrive at for
SN~2009ip must also extend to SN~2010mc and probably to other SNe~IIn
as well.  Margutti et al.\ (2013) underscored this same point.  Below
we suggest that the initially faint percursor 2012a event of SN~2009ip
was actually the SN recombination photosphere, and this would apply to
SN~2010mc as well (i.e. we argue that this initial event was not a
pre-SN outburst in SN~2010mc, but the SN itself).

Figure~\ref{fig:phot} shows the light curve of the 2012 event of
SN~2009ip compared to SN~2010mc, plotted on an absolute magnitude
scale as in Smith et al.\ (2013) but with an extended timescale and
added information for comparison.  As a reference, we have set time =
0 to be the well-defined beginning of the sharp rise to peak
luminosity, as noted earlier (this is {\it not} meant to designate the
time of explosion).  The black dots are the $R$-band photometry of
SN~2009ip, including our new late-time measurements, and the
preliminary estimate quoted by Fraser et al.\ (2013b).  The red dots
are the $R$-band photometry for SN~2010mc from Ofek et al.\ (2013a),
as well as our late-time LBT measurements.  The dashed red line
connects our new SN~2010mc photometry point to the late measurement by
Ofek et al.\ (2013), indicating an approximate decline rate of 0.0012
mag day$^{-1}$, which is obviously much slower than radioactive decay.

We also include, for comparison, the $V$-band {\it Swift}/UVOT
photometry for SN~2009ip as a dotted curve, and the integrated
bolometric (UV/visual/IR) luminosity as a solid black curve, both from
Margutti et al.\ (2013).  The $V$-band photometry fades faster than
$R$-band at late times because the SN color becomes redder, and
because of increasing H$\alpha$ line emission as is typical of
SNe~IIn.  The bolometric luminosity is higher than the $R$-band at
peak because the SN is hot and has a strong UV excess (Margutti et
al.\ 2013), but otherwise the $R$-band magnitude tracks the bolometric
luminosity well.  In particular, at the end of the timespan for which
Margutti et al.\ (2013) provide the bolometric luminosity, the curve
appears to level off, and this is consistent with our last late-time
$R$-band measurement before SN~2009ip became unobservable behind the
Sun (around $t$=100 days).

\begin{figure*}
\includegraphics[width=5.3in]{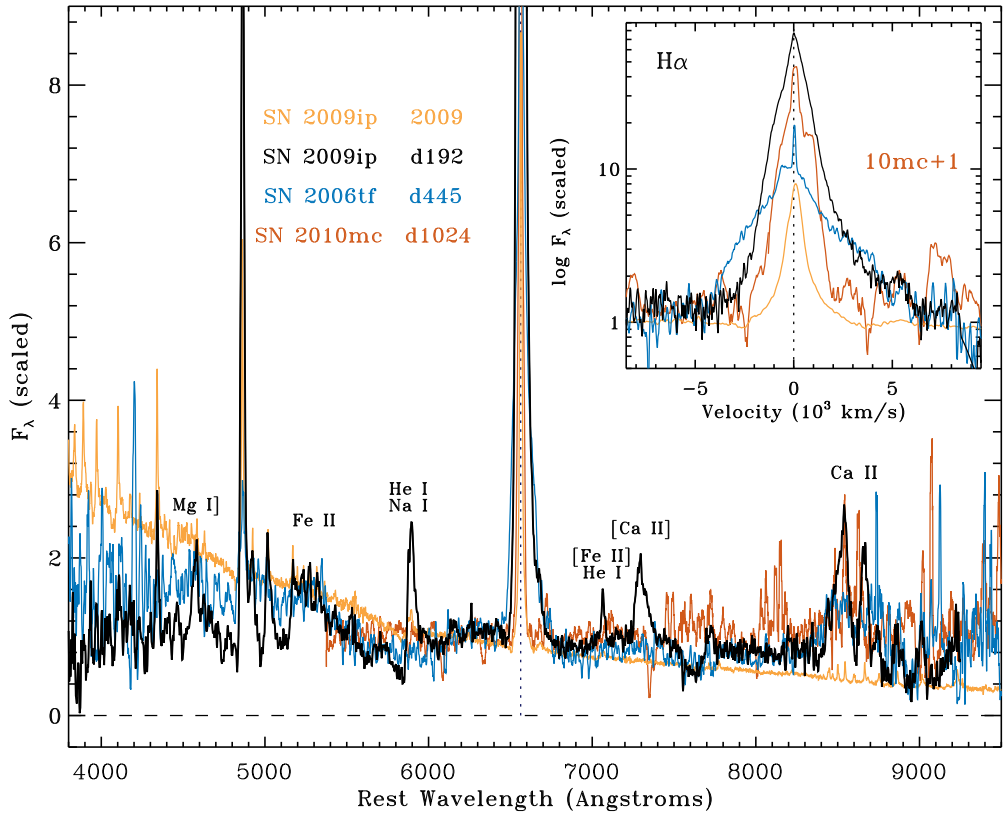}
\caption{A closer look at the late-time spectra, as in
  Figure~\ref{fig:l8spec}, but plotted on a linear intensity scale and
  normalized to the continuum level. Here we compare the day 192
  spectrum of SN~2009ip (black) to the day 445 spectrum of SN~2006tf
  (blue) from Smith et al.\ (2008), and the day 1024 spectrum of
  SN~2010mc (orange).  We also include, for comparison, the pre-SN
  spectrum of the LBV progenitor of SN~2009ip (yellow), taken in
  Sept. 2009, from Smith et al.\ (2010b).  Several likely line
  identifications are listed.  The inset shows the H$\alpha$ line
  profiles on a logarithmic intensity scale, where we have added a
  constant of 1 to the spectrum of SN~2010mc, for comparison to the
  other normalized spectra, since no continuum was detected in this
  spectrum.}
\label{fig:l8spec2}
\end{figure*}

When SN~2009ip reappeared from behind the Sun about 100 days later, it
was roughly 1 mag fainter in $R$.  This rate of fading is consistent
with $^{56}$Co -- $^{56}$Fe radioactive decay luminosity.  This
differs from the claim made by Fraser et al.\ (2013b) that the
late-time fading rate is inconsistent with radioactive decay. For
comparison, Figure~\ref{fig:phot} shows the radioactive decay
luminosity corresponding to a synthesized $^{56}$Ni mass of 0.04
$M_{\odot}$ (dashed gray line), tied to the bolometric luminosity at
$\sim$100 days.  This inferred mass of synthesized $^{56}$Ni depends
on the date that one assumes for the time of the SN explosion
($T_{SN}$).  For reasons outlined below, we set $T_{SN}$ to be just
before the beginning of the 2012a precursor event, at roughly $-$47
days in Figure~\ref{fig:phot}.
%If $T_{SN}$ is actually later, like at the beginning of the 2012b
%event in a separate explosion as advocated by Margutti et al.\ (2013),
%then the maximum mass of $^{56}$Ni implied by the late time decay
%would be lower.

We caution that although the day 100-200 fading rate appears to be
entirely consistent with radioactive decay, we know that the H$\alpha$
luminosity is substantial and that SN~2009ip still has ongoing CSM
interaction that makes a contribution to the $R$-band luminosity.  At
later epochs, the light curve appears to flatten out.  The latest decay
rate appears very similar to that of SN~2010mc, although at a lower
overall luminosity.  This means that radioactive decay contributes a
smaller and smaller fraction of the luminosity as time proceeds, so
the best time to detect a signature of radiactive decay would be days
100-200.

In fact, most SNe~IIn do seem to have light curves that flatten at
late times due to ongoing CSM interaction, although a wide variety of
decline rates are seen at different times, due to different CSM
density and radial extent.  For example, SN~2005ip showed a late-time
luminosity that was essentially flat for several years (Smith et al.\
2009b; Fox et al.\ 2011; Stritzinger et al.\ 2012), whereas the
decline rate for SN~1998S measured by Li et al. (2002) was much
steeper, although not as steep as for pure radiaoctive decay.  Indeed,
at very late times (5000+ days) SN~1998S did eventually flatten-out as
well, as indicated by measurements published recently (Mauerhan \&
Smith 2012).  A spectrum at that time showed very clear signs of CSM
interaction in SN~1998S, with some fast and oxygen-rich material still
crashing into the reverse shock, and was not consistent with a return
to a surviving luminous star.  The possibility that oxygen-rich ejecta
may take several years to cross the reverse shock is relevant to the
discussion of the ``nebular phase'' spectra below (Section 4.2).  Our
measurement of the late-time decline rate of SN~2010mc shows that it
is intermediate between the normal SN~IIn 1998S and the unusually
long-lasting SN~2005ip, and its late-time spectrum is consistent with
ongoing CSM interaction (see below).  In the originally submitted
version of this paper, we stated that it should not be surprising if
SN~2009ip follows a similar trend in coming years.  Indeed, more
recent data now confirm that SN~2009ip's late-time light curve is
flattening out.  This flattening is still $\sim$100 times more
luminous than the progenitor star, so this does not indicate that a
star survived the event.

What about the main event?  As proposed by Mauerhan et al.\ (2013a),
we have adopted the hypothesis that the initial 2012a brightening of
SN~2009ip was the actual SN explosion event, and that the very broad
Balmer lines seen in the spectrum at that time (Smith \& Mauerhan
2012a; Mauerhan et al.\ 2013a) were produced in the rapidly expanding
SN ejecta recombination photosphere.  Similarly, we have attributed
the 2012b event and its very rapid rise (Prieto et al.\ 2013) to mark
the onset of intense CSM interaction, when some of the very fast
ejecta from the 2012a event crashed into the CSM produced by events
1-3 yr earlier.  The 2012a event is somewhat faint compared to the
peak luminosities of normal Type Ibc and Type II SNe (typically around
$-$16 to $-$17 mag), but it is also true that one expects to see an
initially lower luminosity for a relatively compact massive
progenitor, such as the case of SN~1987A (e.g., Arnett et al.\ 1989).
Indeed, the early-time luminosity of the 2012a event is the same as in
SN~1987A at a similar epoch (Figure~\ref{fig:phot}).  As another
example of a relatively faint SN, Figure~\ref{fig:phot} shows the
$R$/unfiltered light curve of the recent peculiar event SN~2010jp
(Smith et al.\ 2012), as a blue curve that provides a reasonable match
to the luminosity evolution of the 2012a event of SN~2009ip, until the
moment when CSM interaction turns on.  SN~2010jp was thought to be an
unusual jet-driven SN, which may have produced a relatively low mass
of synthesized $^{56}$Ni (about 0.003 $M_{\odot}$), perhaps because of
fallback into the same black hole that launched the jets (Smith et
al.\ 2012).  This is predicted to occur in some massive stars at
relatively low metallicity (e.g., Heger et al.\ 2003).  In that case,
this comparison (Fig.~\ref{fig:phot}) to the initial 2012a peak of
SN~2009ip is potentially quite interesting, since both SN~2009ip and
SN~2010jp were massive stars that exploded in the extremely remote
(and presumably lower-metallicity) regions of their host galaxies.
Indeed, both Margutti et al.\ (2013) and Fraser et al.\ (2013a)
inferred a sub-solar metallicity for the remote location of SN~2009ip.
Moreover, SN~2010mc resides in a rather faint dwarf galaxy host
(Figure~\ref{fig:img}; Ofek et al.\ 2013a), for which we measure an
absolute $r^{\prime}$ magnitude of roughly $-$16.9 mag in our LBT
image; this places its host galaxy at a luminosity comparable to that
of the Small Magellanic Cloud, strongly suggesting sub-solar
metallicity (probably 0.2--0.4 $Z_{\odot}$).  The behavior of the
2012a event is discussed in more detail in Section 4.

One further comparison deserves mention.  Figure~\ref{fig:phot} also
includes the light curve of the prototypical Type IIn SN~1998S, with
$R$-band photometry from Fassia et al.\ (2000) as well as more
recently published very early time photometry on the rise to peak from
Poon et al.\ (2011).  SN~1998S was about 0.5 mag more luminous at
peak; if it is scaled to match the 2012b peak of SN~2009ip, we see
that the shape of the light curve is quite similar.  The very rapid
rise to peak is almost identical to SN~2009ip and SN~2010mc, and the
decay from peak is marginally slower.  At early times near peak in
SN~1998S, the spectrum showed a smooth blue continuum, Lorentzian
profiles of Balmer emission lines, and the Wolf-Rayet ``bump''
(He~{\sc ii} $\lambda$4686 and N~{\sc iii}) in emission (Leonard et
al.\ 2000), just like SN~2009ip at its 2012b peak (Mauerhan et al.\
2013; Figure~\ref{fig:earlyspec}).  This comparison is interesting,
since SN~1998S did show a radioactive decay tail at late times, and it
did eventually show oxygen-rich ejecta crossing the reverse shock at
very late times (Mauerhan \& Smith 2012).  It has also been proposed
that the CSM of SN~1998S was concentrated in an equatorial ring
(Leonard et al.\ 2000), as we suggest for SN~2009ip (see
below).

\subsection{Early Spectral Evolution}

Figure~\ref{fig:earlyspec} compares our spectra of SN~2009ip to the
available spectra of SN~2010mc at early times around peak luminosity,
in chronological order relative to the same initial onset of the rise
to peak.  It is clear from this comparison that the spectra form a
continuous time sequence, and that the nearly identical character of
the light curves extends to their spectral evolution.  One minor
difference is that at all epochs (including at later times), He~{\sc
  i} emission lines tend to be stronger in SN~2009ip.  Since the
apparent continuum temperatures are similar, this may indicate a
higher He abundance in SN~2009ip.

The key characteristics displayed by the early-time spectra are (1) a
hot blue continuum, (2) strong narrow Balmer emission lines with broad
Lorentzian wings, (3) much fainter underlying broad emission and
absorption components that are not always easily seen, and (4) an
emission bump from He~{\sc ii} at the very earliest times.  These
properties are quite commonly seen in other normal SNe~IIn as well,
including prototypical objects like SN~1998S and others (Leonard et
al.\ 2000; Kiewe et al.\ 2012; Gal-Yam et al.\ 2007; Smith et al.\
2008, 2009b, 2010a; Taddia et al.\ 2013).  For comparison,
Figure~\ref{fig:earlyspec} also shows an early-time spectrum of
SN~2006tf, which is a super-luminous SN IIn and quite likely to be a
terminal SN explosion (Smith et al.\ 2008).

As SN~2009ip reached its 2012b peak, the very broad emission and
absorption components that had been present in its spectrum in the
2012a event became less prominent or even seemed to disappear.  This
is thought to be due to the increased continuum luminosity from CSM
interaction causing a photosphere outside the forward shock, and is a
common property of SNe~IIn (although the presence of broad lines {\it
  before} the peak has not actually been observed in other SNe~IIn,
since no other object has such early pre-peak spectra available).  The
underlying broad P Cygni profile of H$\alpha$ is faintly visible in
our day 12 spectrum of SN~2009ip (Figure~\ref{fig:earlyspec}); at this
epoch, we find that we can roughly match the strength and profile of
the observed broad H$\alpha$ component by combining the day $-$17
(2012a) spectrum with a 13,000~K blackbody, contributing roughly 5\%
and 95\% of the continuum flux, respectively. Thus, from the spectrum
it would seem that the underlying broad SN photosphere spectrum never
disappeared completely, but was simply outshined by the very bright
and hot continuum from CSM interaction at the peak of 2012b.  This is
quite consistent with our interpretation of the light curve in
Figure~\ref{fig:phot}, where the underlying SN (assuming that its
intrinsic light curve is similar to SN~2010jp) would be $\sim$3 mag
fainter than the peak, indicating that it contributes less than 10\%
of the observed flux.  Similarly, the underlying broad component of
H$\alpha$ is faint but clearly present in the day 14, 21, and 27
spectra of SN~2010mc (Figure~\ref{fig:earlyspec} and Ofek et al.\
2013a).

\subsection{Late-Time Spectral Evolution}

As SN~2009ip faded from its 2012b peak luminosity, the broad
components in the line profiles reappeared, as shown by Mauerhan et
al.\ (2013a).  This is also commonly seen in normal SNe~IIn, such as
SN~1998S (Leonard et al.\ 2000), SN~2005gl (Gal-Yam et al.\ 2007), and
others.  In CSM interaction, the thin cold dense shell of post-shock
gas continues to expand.  If the density of the pre-shock CSM drops,
then the CSM interaction luminosity also drops.

The freely expanding SN ejecta are seen at late times, as confirmed by
the fact that the underlying broad-line spectrum of SN~2009ip at day
60, excluding the narrow emission-line components, is matched well by
a spectrum of a normal SN~II-P at a comparable epoch.
Figure~\ref{fig:l8spec} compares the day 60 spectrum of SN~2009ip
(black) to a spectrum of SN~1999em (red); here we have combined the
SN~1999em spectrum with a blackbody (each contributes half of the red
continuum flux) to mute the strength of the absorption and emission
lines, as would be expected if some of the continuum luminosity is
still contributed by CSM interaction (this is the case in other
SNe~IIn as well; see see Smith et al.\ 2010a).  The shape of the
continuum, the undulations in the continuum due to blended broad
absorption lines, the broad component of H$\alpha$, and the very broad
P Cygni profile of the Ca~{\sc ii} IR triplet in SN~2009ip are all
reminiscent of a SN~II-P.  Margutti et al.\ (2013) also showed a
similar comparison, noting that SN~2009ip at 51 days after peak
(roughly day 60 relative to the minimum) closely resembled the
spectrum of the SN II-P 2006bp at a similar epoch (day 73).  However,
Margutti et al.\ (2013) chose to interpret SN~2009ip as a non-SN
event.

Similarly, Figure~\ref{fig:l8spec} shows that the day 82 spectrum of
SN~2009ip is very well matched by a day 81 spectrum of SN~1987A during
its decline from peak, also with 50\% of the red continuum contributed
by a blackbody (and of course, narrow emission lines from CSM
interaction are missing in SN~1987A's spectrum).  The broad P Cygni
profile of He~{\sc i} $\lambda$5876 is almost identical in both
objects, whereas the broad P Cygni component of H$\alpha$ has similar
strength but $\sim$10\% lower velocities in SN~1987A.

The fact that we see a SN-like continuum with broad emission lines ---
even at late times after SN~2009ip fades from its peak luminosity ---
is extremely important for understanding the true nature of SN~2009ip.
In order to see this underlying SN continuum with broad P Cygni lines,
the expanding fast ejecta must remain optically thick until late
times.  The relevant timescale is the diffusion time for an expanding
envelope, given by $\tau_{\rm d} \simeq 23$ days ($M/R_{15}$), where
$M$ is the mass of the ejected envelope in $M_{\odot}$ and $R_{15}$ is
the radius in units of 10$^{15}$ cm (see Smith \& McCray 2007; Arnett
1996).  For the SN ejecta to remain opaque at day 60, at which time
Margutti et al.\ (2013) calculate a minimum blackbody radius (assuming
spherical symmetry) of 1.5$\times$10$^{15}$ cm, we would require at
least $\sim$4 $M_{\odot}$ in the expanding stellar envelope.  If the
time of explosion is actually 47 days earlier, as we advocate in this
paper, then the required mass is even larger (at least 6 $M_{\odot}$).
These mass estimates are quite approximate (factor of $\sim$2
uncertainty), but it seems clear that the explosion ejecta must
substantially exceed 1 $M_{\odot}$.  At the same epoch (day 60), we
measure a FWHM for the H$\alpha$ emission line of 8940 ($\pm$100) km
s$^{-1}$.  This speed would imply an explosion energy of roughly
3$\times$10$^{51}$ergs (or more for an explosion date at $-$47 days).

Quoted ejecta mass values of $\sim$0.1 $M_{\odot}$ (Margutti et al.\
2013; Fraser et al.\ 2013a) are estimated by the {\it minimum} mass
needed to account for the peak luminosity through CSM interaction,
whereas we estimate the mass needed to keep the ejecta opaque for
60-100 days.  These two are not really in conflict, since the
$\sim$0.1 $M_{\odot}$ mass estimated from CSM interaction is only a
lower limit --- it is only a minimum because not all the envelope mass
must participate in CSM interaction (only the fastest material reaches
the shock at early times), and because the CSM may only intercept a
small fraction of the solid angle of the ejecta if the CSM is
asymmetric (as it must be; see below).  The envelope mass could be
lower than estimated from the diffusion time if the luminosity at
60-100 days is powered largely by radioactive decay, but that too
would require a core-collapse SN event.

The persistence of these underlying broad P Cygni components at late
times therefore rules out the interpretation of SN~2009ip's 2012 event
as the loss of a mere 0.1 $M_{\odot}$ in a 10$^{50}$ erg explosive
shell ejection event.  Expanding at $\sim$8000 km s$^{-1}$, 0.1
$M_{\odot}$ of ejecta would become fully transparent in only a few
days.  It is much more straightforward to explain SN~2009ip as a true
core-collapse event.

\begin{figure}
\includegraphics[width=3.1in]{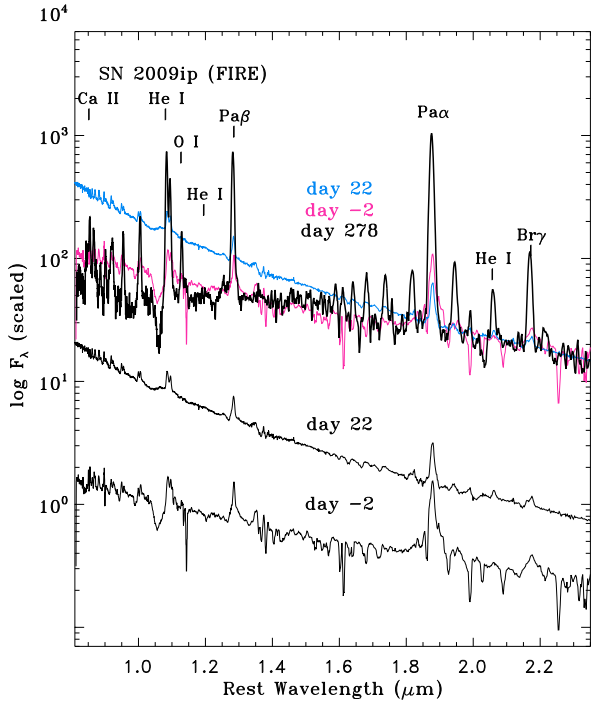}
\caption{A late-time near-IR spectrum of SN~2009ip obtained on day 278
  using FIRE on Magellan, compared to two earlier IR spectra obtained
  with the same instrument taken just before the rise (day -2) and at
  peak luminosity (day 22), from Smith et al.\ (2013).  A FIRE
  spectrum was also obtained on 2013 April 17 (day 206), but it looks
  very similar to the day 278 spectrum, and is not shown here.}
\label{fig:irspec}
\end{figure}

After day 60, the broad emission lines get stronger as the continuum
continues to fade.  Particularly noteworthy is the increasing strength
of the broad emission from the Ca~{\sc ii} IR triplet around 8500 \AA,
which becomes very strong by the time SN~2009ip reappeared from behind
the Sun (after day 190).  This is typical of core-collapse SNe.
SNe~IIn, in particular, tend to have Ca~{\sc ii} emission absent at
early times, whereas at 50-100 days this broad feature appears and
becomes strong (in some cases comparable to H$\alpha$).  Strong and
broad Ca~{\sc ii} emission is not ever seen in LBVs (e.g., Smith et
al. 2011a), at least not at such high speeds of several 10$^3$ km
s$^{-1}$.  In Figure~\ref{fig:l8spec2} we show our late-time spectra
of SN~2009ip on a linear scale, with the flux normalized to the
continuum level; Figure~\ref{fig:l8spec2} shows the pre-SN spectrum of
the LBV progenitor of SN~2009ip, in which the Ca~{\sc ii} IR triplet
is much weaker and narrow, with no P Cygni absorption component.

Figure~\ref{fig:l8spec2} also shows the late-time (day 445) spectrum
of SN~2006tf (Smith et al.\ 2008) for comparison.  Except for the
fainter He~{\sc i} lines and the 7300 \AA \ feature in SN~2006tf
(probably indicative of SN~2009ip's higher He abundance and underlying
nebular spectrum; see Section 4), the spectrum is very similar to the
day 193 spectrum of SN~2009ip, including the flat shape of the
continuum (or psuedo continuum from blended emission lines),
H$\alpha$, and the Ca~{\sc ii} IR triplet.  These late-time spectra
are very unlike the pre-SN spectrum of SN~2009ip, shown (in yellow) in
the same figure with the same red continuum normalization.  The pre-SN
spectrum of SN~2009ip has a much steeper blue continuum, weaker
emission lines, and much narrower emission lines.  The inset of
Figure~\ref{fig:l8spec2} compares the H$\alpha$ profiles of SN~2009ip,
SN~2010mc, and SN~2006tf at late times, along with the H$\alpha$
profile of the SN~2009ip progenitor.  The day 190-400 spectra of
SN~2009ip and SN~2006tf share similar broad profiles that are unlike
the much weaker and narrower Lorentzian profile of the progenitor LBV
star.  We noted in the introduction that although the LBV progenitor
of SN~2009ip did show some high speed material in absorption before
its 2012 event, it had never exhibited the broad Balmer P Cygni
emission lines that are characteristic of SN ejecta photospheres.

Our late-time spectrum of SN~2010mc, taken on day 1024, is plotted as
well in Figure~\ref{fig:l8spec}.  By this very late epoch, we see that
the continuum has continued to fade, so that the spectrum is dominated
by H$\alpha$ emission.  The H$\alpha$ emission has also become
narrower, with a width of $\pm$2000 km s$^{-1}$, presumably because
the fast ejecta have either become transparent or because they have
decelerated in ongoing CSM interaction.  This indicates that even
after $\sim$3 yr, SN~2010mc is still not returning to an LBV-like
state with a strong blue continuum and Lorentzian line profiles.

Finally, the near-IR spectra obtained on days 204 and 278 also show
that the late-time spectra of SN~2009ip have taken on a new character,
unlike spectra seen in previous epochs of this object.  Similar to the
visual-wavelength spectra where the continuum has faded and H$\alpha$
is much stronger than at any previous time, the IR spectrum shows many
of the same emission lines as before, but the lines are relatively
much stronger compared to the continuum (Figure~\ref{fig:irspec}).  In
addition to the lines labeled in Figure~\ref{fig:irspec}, one can
clearly see several lines of the H~{\sc i} Brackett series and Paschen
series in emission. The increasing relative strength of both H and
He~{\sc i} lines is probably due to the explosion becoming optically
thin and the continuum fading, which appears as a large increase in
the emission-line equivalent width for lines formed in the optically
thin SN ejecta and shocked shell.
%Indeed the day 278 IR spectrum in Figure~\ref{fig:irspec} resembles
%the near-IR spectra of typical SNe II-P. 
The evolution of the H$\alpha$ equivalent width is discussed below.

\subsection{Equivalent Width Evolution}

\begin{figure*}
\includegraphics[width=6.3in]{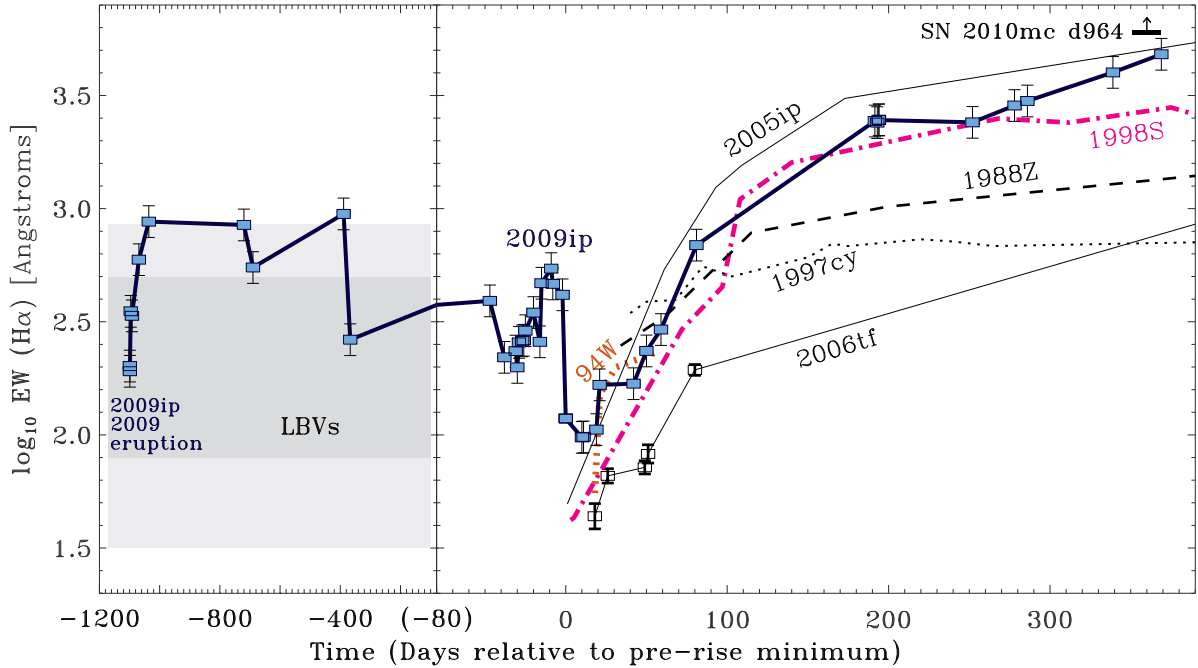}
\caption{The measured H$\alpha$ emission equivalent width of SN~2009ip
  (blue squares) compared to a number of other SNe IIn (the other SNe
  were compiled by Smith et al.\ 2010b; 2008).  At early pre-SN times
  (left panel) we include H$\alpha$ EW measurements from spectra
  kindly provided by A.\ Pastorello (Pastorello et al.\ 2013) combined
  with spectra from Smith et al.\ (2010b).  The gray box indicates the
  typical range of values for the H$\alpha$ EW seen in normal LBVs
  (see Smith et al.\ 2011a).  For SN~2010mc, we plot a lower limit
  (due to an upper limit on the continuum flux) representing the EW on
  day 964, which would be far off the right side of this plot.}
\label{fig:ew}
\end{figure*}

Figure~\ref{fig:ew} shows the total H$\alpha$ emission equivalent
width (EW) for SN~2009ip, for the LBV-like progenitor variability and
for the 2012 event itself (plotted in two sections with different
scales on the time axis, but the same EW scale). This is the total
H$\alpha$ EW, measured as the ratio of the excess emission line flux
to the adjacent continuum flux density at high velocities on either
side of H$\alpha$ (the broad absorption present in the line acts to
slightly weaken the EW measured this way).  We supplemented our pre-SN
spectra with spectra of SN~2009ip from Pastorello et al.\ (2013;
kindly provided by A. Pastorello), some of which are also available on
the WISeREP database.  The dominant source of uncertainty in the EW
measurements comes from the noise in the faint continuum level; we
conservatively adopt error bars of $\pm$20\% in the EW measurements
plotted in Figure~\ref{fig:ew}.

During the $\sim$3 yr of pre-SN evolution, the EW varies but was
typically several 10$^2$ \AA, which is consistent with known LBVs.
The strongest H$\alpha$ EW measured for an LBV is for $\eta$~Car, with
a variable EW of typically 400--800 \AA \ (Smith et al.\ 2003; Stahl
et al.\ 2005).  Other LBVs typically have weaker H$\alpha$ (100-200
\AA) because of their weaker winds compared to $\eta$ Car.

However, starting with the 2012 explosion, the H$\alpha$ EW of
SN~2009ip took on very different and divergent behavior.  At the time
of its rapid brightening just after day 0 in Figure~\ref{fig:phot},
the EW (Fig.~\ref{fig:ew}) dropped substantially due to the added high
continuum luminosity from optically thick CSM interaction.  As the
continuum luminosity faded, however, the H$\alpha$ emission EW rose
sharply to a very high value at late times.  The EW has now risen to
above 4000 \AA, 100 times higher than has ever been seen in any
LBV-like star; this rise is very typical of SNe~IIn (Smith et al.\
2008, 2009b, 2010b).  One reason that the H$\alpha$ EW is so
physically significant is because in an LBV, the H$\alpha$ emission
traces a radiatively driven wind, which in turn requires a strong
UV/optical continuum luminosity source to drive the wind --- on the
other hand, a shock plowing into CSM can produce strong H$\alpha$ but
requires no continuum luminosity to push it.  This is an important
point.

Figure~\ref{fig:ew} shows the time evolution of H$\alpha$ EW for other
classic well-studied SNe~IIn measured the same way, including
SN~1988Z, SN~1998S, SN~1994W, SN~1997cy, SN~2005ip, and SN~2006tf
(these were originally compiled by Smith et al.\ 2008, 2010b).  In
general, SNe that have lower peak luminosities have stronger H$\alpha$
equivalent width, but in all cases shown here of SNe~IIn dominated by
prompt CSM interaction luminosity, the H$\alpha$ EW displays the same
basic rising evolution with time.  Since most SNe~IIn are discovered
near peak luminosity, we cannot compare the pre-SN EW evolution of
SN~2009ip to other SNe IIn --- but from peak luminosity onward,
SN~2009ip is entirely consistent with other normal SNe~IIn.  The EW
evolution of SN~2009ip most closely resembles that of SN~1998S, but is
quite similar to SN~1988Z and SN~2005ip as well.  The rising EW is
understood simply as a result of the CSM interaction region becoming
more optically thin as the shell expands and becomes less dense,
cooler, and more transparent.  This causes the continuum luminosity to
drop while the H$\alpha$ line photons from the shock can more easily
escape (see Smith et al.\ 2008).  For this reason, the increasing
H$\alpha$ EW at late times is generally accompanied by a
transformation from a Lorentzian profile (optically thick to electron
scattering) to a Gaussian or irregularly shaped line profile in
SNe~IIn.  For our day 964, 1024, and 1136 spectra of SN~2010mc, we do
not detect the continuum in our spectra, but measured upper limits to
the continuum level provide a lower limit to the H$\alpha$ EW of
roughly $>$6000 \AA. This is plotted in Figure~\ref{fig:ew}, although
not at the proper time of 964 days since this would be far off the
right of the plot.  The EW lower limit seen in the late-time spectrum
of SN~2010mc is consistent with extrapolating the rate of increase in
the EW of SN~2009ip to much later times.  This makes it seem likely
that the EW of H$\alpha$ in SN~2009ip may continue to rise slowly over
the next few years.

\section{DISCUSSION}

The 2012 explosion of SN~2009ip was undoubtedly one of the most
interesting astronomical transient events in recent years.  There have
now been several very comprehensive studies of the detailed
multiwavelength evolution of SN~2009ip (Mauerhan et al.\ 2013a; Prieto
et al.\ 2013; Pastorello et al.\ 2013; Fraser et al.\ 2013a; Margutti
et al.\ 2013; Ofek et al.\ 2013b; Smith et al.\ 2013; Levesque et al.\
2013), but these very extensive and detailed studies have not yet
converged to agreement in our understanding.  The interpretations and
conclusions of these various studies contradict one another and are
sometimes mutually exclusive in the physical constraints that they
find.  The pre-SN observational record of SN~2009ip is unprecedented,
so this disagreement reflects the fact that we still have much to
learn about the end fates of massive stars.

Some authors have advocated a non-SN explanation for SN~2009ip's 2012
event, proposing that it was either the result of a PPI eruption
(Pastorello et al.\ 2013; Fraser et al.\ 2013a, 2013b) or some other
unknown non-terminal explosive shell-ejection event (Margutti et al.\
2013) with an explosion energy of $\sim$10$^{50}$ erg.  Also, Kashi et
al.\ (2013) propose a that the 2012 event was a binary interaction and
not a SN explosion.  In this paper, we adopt the simple hypothesis
first proposed by Mauerhan et al.\ (2013a), wherein SN~2009ip
2012a/2012b was the true core-collapse explosion of an unstable
LBV-like star.  This simple hypothesis has a few key ingredients:

1.  SN~2009ip's 2012a event was a relatively faint core-collapse SN
initially, which then brightened rapidly in the 2012b event due to CSM
interaction.  We suggest that the SN itself was intrinsically faint
(at first) due to a compact blue progenitor.  This is discussed below.
The main 2012b peak of SN~2009ip was then powered by SN/CSM
interaction, as the SN ejecta caught the dense CSM ejected 1-3 yr
earlier.  In this hypothesis, the total radiated energy can be far
less than 10$^{51}$ ergs (as is the case for all Type~IIn SNe except
for super-luminous SNe~IIn) depending on the mass or aspherical
geometry of the CSM. The only strict requirement is that the total
radiated energy cannot {\it exceed} the explosion energy, but it can
of course be much less than the available kinetic energy.

2.  In our preferred scenario, the pre-SN mass loss of SN~2009ip was
akin to that inferred for other SNe~IIn that require dense CSM, the
main difference being that the luminous precursor variability of
SN~2009ip {\it was actually observed}. Other objects are typically
only discovered near the time of peak luminosity, not 3 years before,
and so they may have suffered similar precursor outbursts that evaded
detection. Erratic variability on short timecales could have a natural
origin in nuclear flashes during O and Ne burning, interactions with a
companion in a binary system (Mauerhan et al.\ 2013a; Soker \& Kashi
2013; Kashi et al.\ 2013) due to an enlarged stellar radius during
these phases, or both.  These are not yet on a firm theoretical
foundation, but the rise of instability in the few years before core
collapse - during O and Ne burning - provides a much more natural
explanation (see Smith \& Arnett 2013) than invoking some other
hypothetical and very extreme instability that occurs when the star is
not yet approaching core collapse.

In the remaining subsections of our paper, we explore various
considerations of the light curve and energy budget, other observed
properties of SNe~IIn, and their consequences for the overall
interpretation of the nature of SN~2009ip. We conclude that the
evidence strongly indicates that SN~2009ip was a true core-collapse SN
with CSM interaction, as is typical of other SNe~IIn.  The main
reasons to doubt a true core collapse expressed so far have been: (1)
the initially faint 2012a event, (2) the CSM interaction energy budget
that does not exceed 10$^{50}$ ergs, and (3) the lack of obvious
nebular-phase oxygen lines indicative of core collapse.  Below we
demonstrate that each of these is invalid as an argument against the
core-collapse hypothesis: the initial faintness of the 2012a is fully
consistent with models of SNe from blue supergiants, various aspects
of the energy budget do in fact require 10$^{51}$ ergs, and late-phase
spectra are indeed comparable to those of SN~1987A.

\begin{figure}
\includegraphics[width=3.1in]{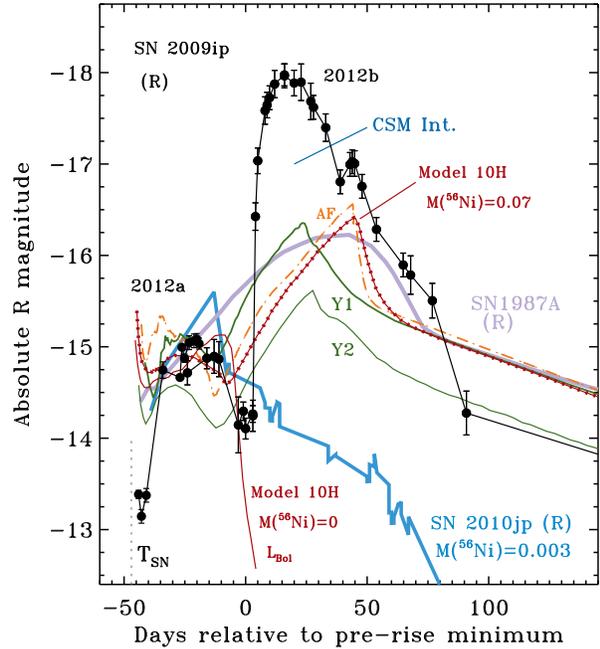}
\caption{A zoomed-in section of the main light curve of the 2012a and
  2012b events of SN~2009ip.  As in Figure~\ref{fig:phot}, SN~2009ip
  is compared to observed $R$-band light curves of SN~2010jp and
  SN~1987A, both of which were initially faint relative to normal SNe
  II-P.  These are compared to model light curves ($L_{Bol}$, not
  $R$-band) from Woosley et al. (1988; red), Arnett \& Fu (1989;
  orange), and Young (2004; green), as discussed in the text.}
\label{fig:models}
\end{figure}

\subsection{The Relatively Faint 2012a Event and the Underlying SN
  Photosphere}

Much of the controversy over the nature of SN~2009ip started during
the 2012a event, when it began to fade after the first report of broad
SN-like emission lines in the spectrum (Smith \& Mauerhan 2012a).  The
fading initially led some (Martin et al. 2012; Margutti et al. 2012a,
2012b) to suggest that it was not a SN after all.  This controversy
continued after the subsequent rapid rise in the 2012b event.  The
light curve of SN~2009ip's ``main event'' is shown in detail in
Figure~\ref{fig:models}, and is discussed below.

Although the initial faintness of SN~2009ip in 2012a (with a peak
absolute magnitude of about $-$15) may have been surprising, it should
not have been.  SN~1987A made it well known that the initial
luminosity of a Type II SN explosion depends primarily on the
progenitor star's radius, and the effect of initial radius on the
light curve has been explored numerically (e.g., Woosley et al.\ 1988;
Arnett \& Fu 1989; Young 2004; Kasen \& Woosley 2009; Dessart et
al. 2013).  With a smaller initial radius, more thermal energy is lost
to adiabatic expansion before the ejecta can radiate.  This is why a
normal SN~II-P from a large red supergiant has a higher initial
plateau luminosity than a SN from a more compact blue supergiant, for
the same explosion energy and H envelope mass.  Since the progenitor
of SN~2009ip was thought to be an LBV-like star with a radius similar
to a blue supergiant, a relatively faint luminosity at early times
should have been expected.

A complicating aspect that may have caused confusion was the initial
fading of SN~2009ip from its 2012a peak (during days $-$20 to 0 in
Figure~\ref{fig:models}), producing a ``valley'' before the rapid rise
to peak in 2012b.  Although the luminosity of SN~1987A was almost
identical to that of SN~2009ip from times soon after explosion to the
2012a peak, this subsequent ``valley'' was not present in SN~1987A.

It is perhaps somewhat ironic or even amusing, then, to recall that
this valley was actually predicted in models for SN~1987A's light
curve --- in fact, the observed absence of the valley was initially
quite puzzling for the case of SN~1987A. In models discussed by
Woosley et al.\ (1988), for example, the valley at 20-40 days after
explosion is a straightforward prediction.  The initially faint peak
is the H recombination plateau due to shock-deposited energy, whereas
a second brighter peak follows when trapped energy deposited by
radiactive decay is able to leak out 1--2 months afterward.  One such
model from Woosley et al.\ (1988) is shown in Figure~\ref{fig:models},
labeled as Model 10H with a synthesized $^{56}$Ni mass of 0.07
$M_{\odot}$.  This model actually gives an excellent match to
SN~2009ip with two exceptions: (1) it does not include CSM interaction
that poweres the main 2012b peak, and (2) it has a higher $^{56}$Ni
mass and more luminous radioactive decay tail than in SN~2009ip.
Maybe it is just a coincidence, but Model 10H from Woosley et al.\
even shows a narrow peak at 45 days ($\sim$90 days after explosion),
at which time SN~2009ip had a small bump during its decline from the
2012b peak.  A similar model from Arnett \& Fu (1989) also shows this
(orange dot-dashed curve in Figure~\ref{fig:models}) in the second
radioactivity peak.  This coincides with the time when strong broad P
Cygni lines returned to SN~2009ip's spectrum (Mauerhan et al.\ 2013a),
suggesting that the underlying SN photosphere was once again
contributing a large fraction of the luminosity.  The reason we are
somewhat hesitant to declare victory in this explanation of the day 45
bump is that the time of this peak in the radioactivity powered light
curve is sensitive to several factors such as the overall $^{56}$Ni
mass, the H envelope mass, explosion energy, mixing, etc.  Different
models with somewhat different parameters (Young 2004), still show
qualitatively the same double-peaked light curve for a blue supergiant
SN, but the time of the radioactive peak is shifted slightly: models
``Y1'' and ``Y2'' in Figure~\ref{fig:models} are very similar but for
a factor of 2 difference in explosion energy and synthesized $^{56}$Ni
mass.  (Note that all the models plotted in Figure~\ref{fig:models}
are for the calculated {\it bolometric} luminosity, whereas the
observed SN light curves are shown as $R$ band.  Thus, the reader
should ignore the very brief few day initial peak following shock
breakout, which is present because of an initially high temperature,
and is not so prominent in red light.)

Why, then, did SN~1987A not have this valley after an initially faint
recombination plateau?  Woosley et al.\ discussed a number of possible
physical mechanisms to erase the valley, but perhaps the most
compelling is mixing (see also Arnett \& Fu 1989).  For a relatively
large synthesized $^{56}$Ni mass, as in SN~1987A, the energy deposited
by radioactivity becomes dynamically important, leading to a density
inversion that may drive Rayleigh-Taylor instabilities in the
expanding ejecta.  This, in turn, will instigate the mixing of
$^{56}$Ni further into the envelope, causing radioactive heating to be
deposited over a larger radius, and to therefore leak out earlier and
smooth-out or even erase the valley.  In effect, mixing causes the
initial recombination plateau and the radioactivity peak to blend
together.  Incorporating this effect, Woosley et al.\ were able to
modify Model 10H to provide a good match to SN~1987A's light curve,
and Arnett \& Fu (1989) achieved similar results with an analytic
light curve.

Judging by the late-time luminosity at 90-200 days, SN~2009ip had a
synthesized $^{56}$Ni mass reduced by about a factor of 2 or more
relative to SN~1987A.  So with all the same explosion parameters but
with roughly 1/2 the $^{56}$Ni mass, we would expect a fainter second
peak powered by radioactive decay, and perhaps also less efficient
mixing of $^{56}$Ni into the envelope, both of which would act to
enhance the valley.  Another influential factor would be the He core
mass --- if SN~2009ip was a more massive progenitor star with a more
massive He core than SN~1987A, the massive He core could decelerate
the expansion of the $^{56}$Ni-rich ejecta, preventing them from being
mixed further out into the H envelope, and thus delaying the rise of
the second radioactivity powered peak.

Continuing with this line of reasoning, for even lower $^{56}$Ni
masses, the second radioactivity powered peak should be even less
luminous, whereas the initial faint recombination plateau could be the
same for a similar H envelope mass and stellar radius.  For only 10\%
of the $^{56}$Ni mass of SN~2009ip, it seems intuitively reasonable to
expect that the second peak might not be a peak at all, but could just
appear as an extension of the plateau or a slowed decline.  This may
be the explanation for SN~2010jp (shown in blue in
Figure~\ref{fig:models}), for which the $^{56}$Ni mass was inferred to
be 0.003 $M_{\odot}$ or less (Smith et al.\ 2012).  For extremely low
or zero $^{56}$Ni mass, the light curve would simply plummet after
$\sim$40 days, although the initial recombination plateau would be the
same if other parameters are held constant.  Another model from
Woosley et al.\ (1988) is shown in Figure~\ref{fig:models}; this is
very similar to the Model 10H discussed above, but with the
synthesized $^{56}$Ni mass artificially set to zero.  Note that this
faint peak is for a true core-collapse SN explosion with 10$^{51}$ of
ejecta kinetic energy, not a weak or failed SN.

Thus, by comparison with published model light curves intended for
SN~1987A, we see that a 10$^{51}$ erg core-collapse explosion of a
blue supergiant gives a self-consistent explanation for the light
curve of SN~2009ip, if we admit that the 2012b peak is primarily
powered by CSM interaction.  This is also consistent with the observed
spectral evolution that exhibited 4 main phases: (1) the initial faint
2012a peak was dominated by broad P Cygni lines characteristic of a
recombination plateau photosphere, (2) the 2012b peak was dominated by
a hot blackbody and narrow emission lines as is typically seen in SNe
IIn with strong CSM interaction, (3) the fading after the 2012b peak
was once again dominated by broad lines from the underlying fast SN
ejecta that are reheated by radioactivity, and (4) the late-phase
luminosity is due to a combination of a fading radioactive decay tail
and lingering CSM interaction.  Of course, superposed narrow nebular
lines are always seen in the spectrum due to photoionized CSM ahead of
the shock.  Comparison with models mentioned above suggest that, for
an explosion kinetic energy of 10$^{51}$ ergs, the luminosity and
duration of the 2012a peak indicate a hydrogen envelope mass of
roughly 10 $M_{\odot}$ and a progenitor radius of
(3-5)$\times$10$^{12}$ cm ($\sim$60 $R_{\odot}$).  This is appropriate
for the inferred LBV-like status of the progenitor (Smith et al.\
2010b).  For example, the stellar radius of $\eta$~Carinae inferred
from models of the present-day spectrum is about 60-100 $R_{\odot}$
(Hillier et al.\ 2001).

One last point to mention is that the similar initial faint peaks of
SN~2009ip and SN~2010jp may have an interesting connection to low
metallicity.  Both objects are consistent with a very similar
explosion from a relatively compact blue supergiant star, except that
SN~2009ip had a larger $^{56}$Ni mass and stronger pre-SN outbursts
that led to stronger CSM interaction and higher luminosity.  SN~2010jp
was a Type~IIn with CSM interaction, but not strong enough to enhance
the luminosity, and it had evidence for a fast bipolar jet that may
have been related to the lower $^{56}$Ni mass if fallback to black
hole was involved.  The possible connection to low metallicity is that
both SNe exploded in extremely remote environments in the far
outskirts of their host spiral galaxies (Mauerhan et al.\ 2013a;
Fraser et al.\ 2013a; Margutti et al.\ 2013; Smith et al.\ 2012).
SN~2010mc was also found in a presumably low-metallicity dwarf galaxy
(Figure~\ref{fig:img}), and SN~1987A exploded in the moderately
metal-poor LMC.  Arnett et al.\ (1989) have reviewed reasons why
explosions of blue supergiant progenitors might be relatively more
common in lower-metallicity regions.

\begin{figure}
\includegraphics[width=3.1in]{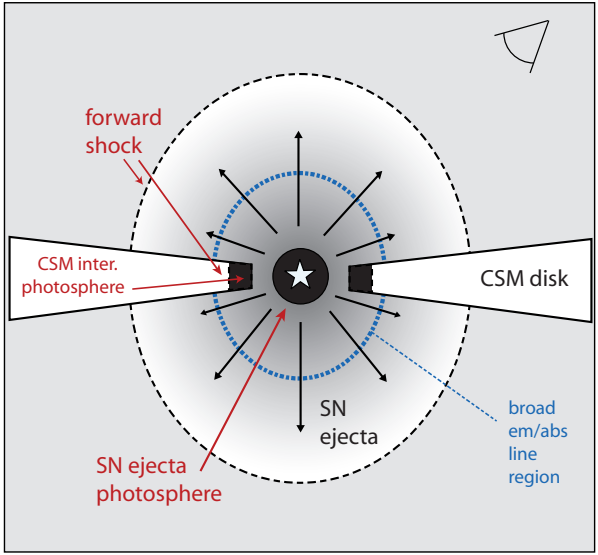}
\caption{A sketch of the likely CSM interaction geometry of SN~2009ip.
  The SN explosion is surrounded by a disk (seen here edge-on), such
  that the SN ejecta (gradient) expand freely in the polar directions
  with the forward shock suffering little deceleration and reaching a
  large radius (and thus, retaining most of their kinetic energy).  In
  the equatorial plane, however, the SN ejecta are stopped as they
  crash into the CSM disk (so that the forward shock reaches a much
  smaller radius), and their kinetic energy is converted into
  radiation.  There is a normal SN photosphere receding into the inner
  SN ejecta, but there is also a photosphere formed in the optically
  thick CSM-interaction region in the equatorial plane, which
  dominates the luminosity at the peak of SN~2009ip.  The blue dashed
  line represents the outer extent of the region where broad emission
  and absorption lines arise in the SN ejecta.  The symbol in the
  upper right denotes a likely intermediate-latitude observer's
  viewing angle.}
\label{fig:geom}
\end{figure}

\subsection{CSM Interaction and the Energy Budget}

Another key component in the debate about the nature of SN~2009ip
(true SN or not?) is that, in principle, a high explosion energy of
10$^{51}$ erg is not necessarily required to produce a bright SN-like
display.  This is because CSM interaction can be an extremely
efficient engine for converting ejecta kinetic energy into radiated
energy; this same mechanism allows super-luminous SNe to radiate 100
times more light than a normal SN with the same explosion kinetic
energy (e.g., Smith \& McCray).  Both Woosley et al.\ (2007) and
Dessart et al.\ (2010) have pointed out that a true core-collapse SN
is not necessarily required to produce a SN-like transient; even
explosions of 10$^{50}$ ergs can achieve peak luminosities like SNe
without any radioactive decay luminosity.  A similar line of reasoning
allows a low-energy electron-capture explosion to produce a bright SN,
as in the case of the Crab Nebula and SN~1054 (Smith 2013b).  This
same argument has been extended to SN~2009ip's 2012 event, advocating
that SN~2009ip was a PPI ejection or some other non-SN event
(Pastorello et al.\ 2013; Fraser et al.\ 2013a; Margutti et al.\
2013).  These authors find that only 10$^{50}$ erg is required to
produce the observed radiated energy, not 10$^{51}$ ergs.  This is,
however, only part of the story.

Two points about the energy budget are essential to mention: First, a
low radiated energy of only $\sim$3$\times$10$^{49}$ ergs (Margutti et
al.\ 2013) does not provide evidence against the SN hypothesis.
Normal core-collapse SNe typically radiate only 10$^{49}$ ergs or less
(Arnett 1996).  This is because most of the 10$^{51}$ ergs of
explosion energy (excluding neutrinos) gets converted to kinetic
energy through adiabatic expansion, which is even more severe for
relatively compact blue supergiants.  Only through CSM interaction
does some fraction of this energy get converted back to radiation.
Second, even with intense CSM interaction, the total radiated energy
can still be far less than the ejecta kinetic energy.  The explosion
kinetic energy required to power CSM interaction is merely {\it a
  lower limit}, involving only the portion of the ejecta that
participate in the shock interaction.  The efficiency of converting
kinetic energy into radiation can be high (i.e. 50\% or more) if
conditions are right (high mass and slow CSM, fast SN ejecta,
spherical symmetry), but the efficiency can be far less.

Asymmetry is an essential consideration (perhaps the most important
one) in the energy budget from CSM interaction (see
Figure~\ref{fig:geom}).  We have mentioned the persistent broad lines
in the spectrum earlier in this paper in the context of the long
diffusion time and high optical depths in high speed ejecta at late
times, basically ruling out a low-mass/low-energy shell ejection
because a 10$^{51}$ erg explosion is required by a few $M_{\odot}$
moving at 8000 km s$^{-1}$.  The persistence of these broad lines is
also directly relevant to the asymmetry of the explosion and the
energy budget.  Since we still see broad H$\alpha$ emission lines with
widths of 8000 km s$^{-1}$, long after the most intense phase of CSM
interaction has ended, the ejecta were not decelerated --- thus,
significant large-scale asymmetry is a strict requirement.  If
SN~2009ip were spherically symmetric, any ejecta faster than
$\sim$2000 km s$^{-1}$ would have crashed into the reverse shock
during the peak of the 2012b event.  In other words, at day 60, the
8000 km s$^{-1}$ ejecta should reach $R$=4$\times$10$^{15}$ cm, larger
than the effective blackbody radius of $R_{BB}$=1.5$\times$10$^{15}$
cm deduced by Margutti et al.\ (2013) around this time assuming
spherical symmetry. The broad lines persist until late times, so a
substantial fraction of the fast ejecta must never interact with the
slow ($\sim$600 km s$^{-1}$) CSM, requiring that large sections of the
explosion's solid angle do not participate in CSM interaction during
the 2012b peak.  Since 10$^{50}$ ergs was the amount of kinetic energy
required to fuel the {\it observed} CSM interaction luminosity, the
true explosion kinetic energy must be substantially higher.  If the
CSM is located in a thin disk or ring, as proposed by some authors for
a number of other SNe~IIn like SN~1998S (Leonard et al.\ 2000) and
SN~1997eg (Hoffman et al.\ 2008), then the true explosion energy could
easily be 10 times higher than the 10$^{50}$ ergs inferred from the
minimum requirements of CSM interaction.  For example, if the CSM is
in a disk with an opening angle of 10$^{\circ}$ (i.e. $\pm$5$^{\circ}$
from the equatorial plane), then this disk will intercept less than
9\% of the SN ejecta's solid angle.  More than 90\% of the fast SN
ejecta will expand unhindered, and will simply cool and fade without
any mechanism to reheat them (Figure~\ref{fig:geom}).

In fact, several authors have inferred a significantly asymmetric CSM
environment for SN~2009ip, including an equatorial disk-like
distribution of the CSM (Levesque et al.\ 2013; Margutti et al.\
2013).  No observations of the polarization of SN~2009ip have been
published yet, but these may prove to be quite interesting to test the
strong asymmetry we infer for CSM-interaction in the 2012b event.

One could imagine that the asymmetry of the CSM is one of the driving
agents to explain the overall diversity among SNe~IIn.  The highest
luminosity events like the super-luminous SNe require a large CSM mass
to decelerate the SN ejecta, but to achieve the radiated energy budget
without appealing to extraordinarily energetic explosions they also
require that the CSM is opaque over most of the solid angle
surrounding the explosion.  SNe~IIn of more moderate luminosity can
result from the same explosion energy if the CSM intercepts a smaller
fraction of the solid angle of the explosion, converting a smaller
fraction of their available kinetic energy into radiation at early
times (Figure~\ref{fig:geom}).  This is consistent with the fact that
many of the moderate-luminosity SNe~IIn also show an underlying
broad-line component in their spectra, indicating that some of the SN
ejecta expand without colliding with the slow CSM.

\subsection{CSM Interaction and the Lack of a Nebular Phase}

A third argument put forward in favor of a non-SN interpretation for
SN~2009ip was that its late-time spectrum does not show a nebular
phase like normal core-collapse SNe, and especially that it does not
show the bright oxygen lines (i.e. [O~{\sc i}]
$\lambda\lambda$6300,6464) seen in other SNe (Fraser et al.\ 2013a,
2013b).  This disregards the fact that the late-time emission of
SNe~IIn is often dominated by CSM interaction, and not by the usual
nebular phase powered by the slow leakage of radioactive energy.  It
also ignores possible complications due to lower $^{56}$Ni mass and
external irradiation of the SN ejecta by CSM interaction luminosity.
In fact, SNe~IIn rarely exhibit a classical nebular phase with a clear
radioactive decay tail because it is masked by strong ongoing CSM
interaction.

Fraser et al.\ (2013a) also claim that SN~2009ip fades at a rate
inconsistent with radioactive decay, and they find no significant IR
excess from dust.  Both of these are in disagreement with observations
reported by us and by other authors (Margutti et al.\ 2013; Smith et
al.\ 2013; this work).  Within observational uncertainties, SN~2009ip
faded at a rate entirely consistent with radioactive decay during days
100-200 (Figures~\ref{fig:phot} and \ref{fig:models}).  From this we
estimate an upper limit of 0.04 $M_{\odot}$ for the mass of
synthesized $^{56}$Ni.  The $^{56}$Ni mass may, however, be even lower
if more of the luminosity comes from CSM interaction, or it may be
somewhat higher if dust causes extinction.

The light curve appears to flatten after day 200, fading more slowly
than radioactive decay as the emergent light is once again dominated
by ongoing CSM interaction, as in most SNe~IIn.  Therefore, the most
promising time to see evidence of a radioactivity-powered nebular
phase is during days 100-200.  After this, the nebular spectrum becomes
more and more difficult to see, as it contributes an ever decreasing
fraction of the observed luminosity.

It is therefore quite interesting that our day 190-200 spectra of
SN~2009ip are matched surprisingly well by observed spectra of
SN~1987A at around the same time (day 152), neglecting the stronger
narrow components in SN~2009ip that are due to external CSM
interaction.  Figure~\ref{fig:l8spec} includes a direct comparison of
late-time spectra of SN~2009ip and SN~1987A.  At this epoch, the
overall shape of the continuum and the strength of the broad lines of
He~{\sc i}, H$\alpha$, and Ca~{\sc ii} are similar in both objects.
More to the point, we note that the broad line around 7300 \AA \ is
indeed clearly detected in SN~2009ip, and has the same strength and
line profile in both objects; this is probably [Ca~{\sc ii}] commonly
seen in nebular spectra of SNe.  It is true that [O~{\sc i}]
$\lambda\lambda$6300,6364 does not appear to be detected in our day
196 spectrum of SN~2009ip, but this line is not very strong in
SN~1987A either (Figure~\ref{fig:l8spec}).  Also, the [O~{\sc i}] line
strength can vary from one core-collapse SN to the next; in SN~1999em,
for example, [O~{\sc i}] $\lambda$6300 was very weak in the day 316
spectrum (Leonard et al.\ 2002), with a strength that would not be
detected in our spectra of SN~2009ip.  Moreover, the first ionization
potential of oxygen is the same as hydrogen.  Since CSM interaction
gives rise to extremely strong H$\alpha$ emission, it may be possible
that Lyman continuum photons generated in the CSM interaction shock
may propagate back into the SN ejecta and ionize oxygen atoms, perhaps
quashing the strength of [O~{\sc i}] lines.  This last idea is
speculative in the absence of detailed radiative transfer
calculations, but the point is that we should not necessarily expect
the nebular phase of a SN~IIn to have exactly the same properties as
normal SNe II-P, where the only source of heating is internal
radioactive decay.  Fraser et al.\ (2013b) claimed that the lack of
evidence for nucleosynthetic products in the late spectrum of
SN~2009ip argues against a core-collapse interpretation, but the
comparison to SN~1987A contradicts this claim.  Aside from [O~{\sc
  i}], which may be affected by photoionization, there is no other
strong line in the spectrum of SN~1987A that should have been detected
in SN~2009ip at a similar epoch (i.e. except for [O~{\sc i}], the
SN~1987A spectrum is always at or below the observed day 193 spectrum
of SN~2009ip in Figure~\ref{fig:l8spec}).

At even later times, it becomes increasingly difficult to see clear
signatures of the nebular phase in spectra.  Figure~\ref{fig:l8spec}
also compares the day 286 and day 339 spectra of SN~2009ip to a
similar epoch (day 293) of SN~1987A.  The spectrum of SN~2009ip at
this late phase is more strongly influenced by CSM interaction,
judging by the blue pseudo continuum and stronger narrow emission
lines that are present in late-phase spectra of well-studied SNe~IIn
like SN~1988Z (Chugai \& Danziger 1994) and SN~2005ip (Smith et al.\
2009b), as well as in a late-time spectrum of SN~2006tf (also plotted
in Figure~\ref{fig:l8spec}). Since the strong 7300 \AA\ feature is
completely absent in the CSM-interation dominated spectrum of
SN~2006tf on day 445, we may infer that its strong presence in
SN~2009ip (and SN~1987A) is primarily due to the nebular-phase
emission from the inner ejecta.  The strength and profile of this
line, as well as the Ca~{\sc ii} IR triplet and the underlying broad
wings of H$\alpha$, are still similar in SN~2009ip and SN~1987A in
these latest epochs.  The [O~{\sc i}] emission is still missing in
SN~2009ip, perhaps for the reasons noted above, but it only needs to
be weaker by a factor of 2-3 to go undetected in SN~2009ip.  As noted
above, however, this [O~{\sc i}] line is similarly weak in late
spectra of the classic SN~II-P 1999em.  If the fading rate of
SN~2009ip continues at its present level, we would expect that the
underlying nebular lines would become progressively more difficult to
see, and that the spectrum will more closely resemble the late
spectrum of SN~2006tf.

One future prospect is that lines normally seen in nebular phases
(like oxygen lines) may become observable again in the future if/when
the oxygen-rich inner layers crash into the reverse shock.  This may
take several years, but predicting when it will occur is difficult.
An unusually massive He core (if SN~2009ip had a very massive
progenitor star) could decelerate the expansion of inner layers that
are rich in oxygen and other nucleosynthetic products, potentially
delaying the time when they finally participate in CSM interaction.
While these O lines are seen at very late times in some SNe~IIn, such
as SN~1998S (Mauerhan \& Smith 2012), other well-studied SNe~IIn have
not shown them (i.e., SN~2006tf, SN~2006gy, SN~2005ip; Smith et al.\
2008, 2009b, 2010a).

\subsection{The Broader Class of Type IIn Supernovae}

In comparing SN~2009ip to other examples of SN explosions, and
especially to the wider class of Type~IIn SNe, it is essential to
remember that the quality and quantity of pre-SN observations for this
source are unprecedented.  Although the behavior of SN~2009ip in its
2012a event seems to be unusual at first glance, we cannot claim with
any confidence that this is unique or unusual behavior among SNe~IIn.
Except for pre-peak photometry of SN~2010mc (no early spectra are
available), this information does not exist for other SNe~IIn.

In fact, all SNe~IIn require some sort of episodic pre-SN mass loss.
For any SN~IIn, the CSM mass that must be ejected in a short amount of
time would require extremely high mass-loss rates that cannot be
provided by normal stellar winds (Smith \& Owocki 2006).  The least
extreme mass loss among SNe IIn progenitors is comparable to the
episodic winds seen in only the most extreme red supergiants like VY
CMa, with mass-loss rates of order 10$^{-3}$ $M_{\odot}$ yr$^{-1}$
(Smith et al.\ 2009a, 2009b).  Normal SNe~IIn require mass-loss rates
of order 0.1 $M_{\odot}$ yr$^{-1}$ (Kiewe et al.\ 2010; Gal-Yam et
al.\ 2007; Chugai et al.\ 2004), only seen previously in giant
eruptions of LBVs, while the more luminous SNe~IIn require mass loss
events comparable to $\eta$ Car's extreme 19th century eruption (Smith
\& McCray 2007; Smith et al.\ 2010b).  To make a SN~IIn at times soon
after explosion, this dense CSM must be ejected in a time period of
only a few years before core collapse.  Thus, all SNe~IIn must do {\it
  something} akin to what was observed in SN~2009ip.

The reason that SN~2009ip appears to be alone (or almost alone;
SN~2010mc) probably has more to do with detection limits than
SN~2009ip being physically unusual.  Recall that most SNe~IIn are
discovered near the time of maximum luminosity or shortly afterward,
using relatively small telescopes that would not be sensitive to
LBV-like eruption luminosities at distances larger than about 30 Mpc.
If we restrict ourselves to a comparison of the data from the time of
SN~2009ip's peak luminosity onward, then there is no indication that
SN~2009ip is significantly divergent compared to the existing sample
of SNe~IIn with similar peak luminosities.  LSST will likely provide
much more valuable constraints on the variability of SN~IIn
progenitors.

\subsection{SN~2010mc continues to fade}

Regardless of the larger class of SNe~IIn, it is clear that SN~2009ip
and SN~2010mc are near twins, at least during their early-time
evolution.  Therefore, with a longer time baseline since its
explosion, SN~2010mc may provide us with a preview of coming
attractions for SN~2009ip.  Basically, in the 2-3 years after its SN
event, SN~2010mc has exhibited no resurgence of erratic LBV-like
variability, it has not had another PPI eruption, and it has not
returned to a luminous LBV-like state as we expect for a surviving
star.  Instead, it has continued to fade and exhibits no detectable
continuum; the late-time spectrum shows only H$\alpha$ emission,
consistent with a shock running into more distant CSM as in other old
SNe~IIn.  The H$\alpha$ equivalent width is over 6000 \AA \ at the
latest phases, more than 2 orders of magnitude stronger than in any
stellar wind.  Its level of late-time CSM interaction appears to be
typical of known core-collapse SNe IIn, comfortably intermediate
between SN~1998S and SN~2005ip (Figure~\ref{fig:phot}).

\subsection{Pulsational Pair Instability Eruptions Are Too Rare}

To provide a powerful explosion that could be as luminous as a typical
SN, the only non-terminal alternative to a real SN proposed so far is
the PPI (e.g., Heger \& Woosley 2002; Woosley et al.\
2007).\footnote{An electron-capture SN can also provide a 10$^{50}$
  erg explosion, and these can produce luminous transients with peak
  luminosity comparable to bright SNe if they have CSM interaction
  (Smith 2013b; Mauerhan et al.\ 2013b), but this is a true terminal
  SN event that undergoes core collapse to yield a neutron star.}
However, the PPI does not provide a very suitable explanation for
SN~2009ip, and is unsatisfactory in the sense that its low rate of
occurance would require SN~2009ip to be different from all other
SNe~IIn.

%%%%%%%%%%%%%%%%%%%%%%%%%%%

There are a number of problems with invoking the PPI to explain
SN~2009ip.  As noted by Ofek et al.\ (2013a, 2013b), the PPI predicts
a much larger mass of ejecta (several $M_{\odot}$) than has been
inferred for the precursor events or the main 2012 event of SN~2009ip
($\sim$0.1 $M_{\odot}$) in the non-terminal CSM interaction models
proposed (Pastorello et al.\ 2013; Fraser et al.\ 2013a; Margutti et
al.\ 2013).  Moreover, as already mentioned by Margutti et al.\
(2013), the PPI predicts a very wide range of timescales for the
pre-SN pulses.  In order for SN~2009ip and SN~2010mc to have identical
$\sim$40 day timescales within the wide range of possible timescales
would require rather extreme fine tuning, making this hypothesis seem
highly improbable.\footnote{Somewhat ironically, the star should
  probably be dead even in the non-terminal PPI hypothesis.  A
  timescale of only 40 days is rather short for the PPI
  instability. According to Heger \& Woosley (2002) and Woosley et
  al.\ (2007), the time between PPI pulses accelerates as the star
  approaches its final core collapse, and the pulse amplitude
  increases. Thus, the observed quickening of the timescale from
  $\sim$1 yr to $\sim$40 days might imply that even if the shell
  ejections were caused by PPI, that the star would have collapsed to
  a black hole by now anyway.  Thus, the PPI leaves us with a scenario
  that is difficult to distinguish from a true SN, since both would
  leave behind a compact corpse and no star.  The only significant
  predicted difference between a true SN and the PPI would be
  $^{56}$Ni production (which should be absent in the PPI scenario),
  but this is also difficult to test in SN~IIn due to late-time CSM
  interaction.}

%%%%%%%%%%%%%%%%%%%%%%%%%%

Another argument against the PPI being the explanation for SN~2009ip
is based on statistics.  Type~IIn SNe comprise 8-9\% of all observed
core-collapse SNe occurring in spiral galaxies (Smith et al.\ 2011b).
Some of these may actually be SNe~Ia with CSM (see e.g., Silverman et
al.\ 2013) and some may be ecSNe (Mauerhan et al.\ 2013b; Smith
2013b).  Even so, a substantial fraction of SNe~IIn are ``normal''
SNe~IIn from massive LBV-like stars or extreme red supergiants.  The
PPI cannot explain the class of SNe~IIn as a whole (or even half of
it) because the PPI occurs in a narrow mass range for very rare, very
massive stars, such that PPI events must be far too infrequent to
explain observed SNe~IIn.  For a normal Salpeter initial mass function
(IMF), the PPI mass range (initial masses of roughly 100-130
$M_{\odot}$; Heger \& Woosley 2002) would contribute less than about
1\% of all exploding stars above 8.5 $M_{\odot}$ (see Smith et al.\
2011b).  Statistics aside, it seems unlikely that the detected
progenitor of SN~2009ip is luminous enough to be in this mass range
(Smith et al.\ 2010b; Foley et al.\ 2011; Margutti et al.\ 2013).  To
make matters worse, though, this quoted PPI mass range is for zero
metallicity; for nearly solar metallicity with mass loss, the mass
range would need to move to higher (and therefore rarer) initial
masses (see, however, Chatzopoulos \& Wheeler 2012).  To propose that
SN~2009ip is not a SN and instead a PPI eruption would require that it
be unique among the rest of the nearby SN~IIn population, which is
currently an unjustifiable claim based on available data.

So, if SN~2009ip wasn't a PPI eruption, then what was it?  What other
mechanism can produce significant eruptive mass loss that precedes an
even more extreme non-terminal explosive event?  No other non-terminal
mechanism has yet been identified.  Without an answer to this, there
is little motivation to further consider a non-SN hypothesis for
SN~2009ip.

\subsection{Final Stages of Nuclear Burning}

The nearly identical light curves of SN~2009ip and SN~2010mc demand an
answer to the question of why these two different objects would have
such similar time evolution.  This makes it seem unlikely that a
binary merger is the explanation (Soker \& Kashi 2013; Kashi et al.\
2013), since both SN~2010mc and SN~2009ip need to be tuned to the same
orbital period.  Margutti et al.\ (2013) highlighted the fact that
both SN~2009ip and SN~2010mc share a $\sim$40 day delay between the
two outbursts, and that this must be an important clue to their
underlying nature, but those authors left the question unanswered.
Indeed, under the hypothesis that the 2012a and 2012b events are two
separate and sequential non-SN explosions (Margutti et al.\ 2013),
this identical timescale in two different objects would require
extreme and unlikely fine tuning, as those authors noted.

However, under the hypothesis we advocate, wherein the 2012a event was
the true SN explosion and the 2012b event was CSM interaction
(Mauerhan et al.\ 2013a), the timescale of 40 days is simply the
amount of time required for the fast SN ejecta to catch up to the
expanding CSM shell ejected 1-2 yr earlier.  If the CSM was ejected at
a speed of $\sim$1000 km s$^{-1}$ and the SN ejecta expand 10 times
faster at 10,000 km s$^{-1}$, then the time needed for the fast SN
ejecta to catch the slower CSM ejected $\sim$1 yr earlier is simply 1
yr $\times$ ($V_{CSM}$/$V_{SN}$) $\approx$ 37 days.  This is roughly
the correct delay time.  In that case, the similar $\sim$40 day delay
between the two sequential outbursts in both SN~2009ip and SN~2010mc
is set by the $\sim$1 yr timescale of the precursor mass ejections.
Thus, we seek a common origin for substantial changes in stellar
structure on timescales of $\sim$1 yr before a powerful explosion.

The very last phases of pre-SN nuclear burning in the core of a
massive star may provide a natural explanation for the erratic
variability of SN~2009ip in the few years leading up to its 2012
demise. In particular, O and Ne burning each last for a time of order
1 yr.  C burning occurs on a timescale that is too long ($\sim$10$^3$
yr) for SN~2009ip's behavior, and Si burning is too quick
(days). Quataert \& Shiode (2012) have suggested that the rapid
burning rate during O and Ne burning and a star's response to extreme
neutrino losses may excite waves that propagate through the envelope,
perhaps leading to enhanced mass loss on a short timescale before core
collapse.  Alternatively, multi-dimensional hydrodynamic simulations
of the shell burning layers during O burning and later phases reveal
potential eruptive instabilities (Meakin \& Arnett 2007; Arnett \&
Meakin 2011).  These instabilities might also help instigate mixing
and perhaps explosive burning, depositing energy into the core that
might lead to hydrodynamic mass ejection (e.g., Dessart et al.\ 2010;
Smith \& Arnett 2013).  Although these mechanisms require a great deal
of additional study, the possibility of disruptive instabilities
arising on the correct timescale before core collapse is intriguing
(see Smith \& Arnett 2013 for a more extensive discussion, including
the possible outcomes in a close binary system).

%Some SNe~IIn do show evidence for strong CSM interaction continuing
%for years, such as SN~1988Z (Chugai \& Danziger 1994), SN~2003ma (Rest
%et al.\ 2011), and more recently SN~2010jl.  This long-lasting CSM
%interaction requires dense CSM over a large range of radii, which in
%turn implies very strong mass loss for centuries before core-collapse.
%The mass-loss rates required for these progenitor stars are also much
%larger than can be provided by normal winds of evolved massive stars.
%Thus, enhanced mass loss arising from instabilities that occur during
%the $\sim$10$^3$ yr of C burning may be relevant in these sources.

In the alternative scenario where SN~2009ip is not a core-collapse SN,
on the other hand, there is no clear physical motivation for the rapid
and accelerating evolution of the instability on timescales of only 1
yr. It is also unclear why these would immediately precede a much more
catastrophic (but also non-terminal) event that involves an explosion
with a significant fraction of the core binding energy.

\subsection{No other SN II\lowercase{n} in history has yet come back to life}

Finally, it is worth recalling an obvious fact.  In the past several
decades, astronomers have found dozens of SNe~IIn at relatively nearby
distances (within $\sim$50 Mpc), and a number of other unusual SN-like
transients.  So far, among objects with a peak absolute magnitude
brighter than $-$16, not one in history has ever had a resurgence as
another violent explosion.  This contradicts expectations of the PPI
hypothesis.  The simplest and most conservative conjecture is that
these stars have indeed met their demise in a final SN event, at least
until strong evidence to the contrary surfaces.

\section*{Acknowledgements}

\scriptsize 

We thank an anonymous referee for a careful reading of the manuscript.
We thank N.\ Morrell for assistance in obtaining some of the data from
Las Campanas, and we thank A.\ Pastorello for sharing spectroscopic
observations of SN~2009ip that contributed to our measurements of the
equivalent width in Figure~\ref{fig:ew}.  JLP thanks support from a
Carnegie-Princeton postdoctoral fellowship.  We thank the staff at the
MMT, LBT, and Las Campanas Observatories for their assistance with the
observations.  Some observations reported here were obtained at the
MMT Observatory, a joint facility of the University of Arizona and the
Smithsonian Institution.  This research was based in part on
observations made with the LBT.  The LBT is an international
collaboration among institutions in the United States, Italy and
Germany. The LBT Corporation partners are: the University of Arizona
on behalf of the Arizona university system; the Istituto Nazionale di
Astrofisica, Italy; the LBT Beteiligungsgesellschaft, Germany,
representing the Max-Planck Society, the Astrophysical Institute
Potsdam and Heidelberg University; the Ohio State University and the
Research Corporation, on behalf of the University of Notre Dame,
University of Minnesota and University of Virginia.

%The supernova research of A.V.F.'s group at U.C. Berkeley is supported
%by Gary \& Cynthia Bengier, the Richard \& Rhoda Goldman Fund, the
%Christopher R. Redlich Fund, the TABASGO Foundation, NSF grants
%AST-0908886 and AST-1211916, and NASA/{\it HST} grants AR-12126 and
%AR-12623 from the Space Telescope Science Institute (which is operated
%by Associated Universities for Research in Astronomy, Inc., under NASA
%contract NAS 5-26555).
%KAIT and its ongoing operation were made possible by donations from
%Sun Microsystems, Inc., the Hewlett-Packard Company, AutoScope
%Corporation, Lick Observatory, the NSF, the University of California,
%the Sylvia \& Jim Katzman Foundation and the TABASGO Foundation.

\end{document}